\documentclass[fleqn,usenatbib]{mnras}
\pdfoutput=1 
\usepackage{newtxtext,newtxmath}

\usepackage[T1]{fontenc}
\usepackage{caption}
\usepackage[font=small,labelfont=bf]{caption}

\usepackage{graphicx}	
\usepackage{color}
\usepackage{floatrow}
\usepackage{cancel} 
\usepackage{array}
\usepackage{amsmath}

\usepackage{booktabs}

\definecolor{purple}{rgb}{0.78,0.18,0.77}

\newcommand{\hmpc}{h^{-1}{\rm Mpc}}
\newcommand{\ihmpc}{{h} \rm{Mpc}^{-1}}

\newcommand{\eq}[2]{%
\begin{equation}\label{#1}
\begin{aligned}
#2
\end{aligned}
\end{equation}
}

\newcommand{\vct}[1]{\boldsymbol{#1}}

\title[Modelling redshift space $k \text{NN-CDFs}$]%
{Modelling $k$NN-CDFs of biased tracers in redshift space using hybrid Lagrangian bias expansion}

\author[Chand et al.]{
Eishica Chand$^{1}$\thanks{E-mail: {\tt eishica.chand@students.iiserpune.ac.in}},
Arka Banerjee$^{1}$\thanks{E-mail: {\tt arka@iiserpune.ac.in}} \\
$^{1}$Department of Physics, Indian Institute of Science Education and Research,
Homi Bhabha Road, Pashan, Pune 411008, India
}

\date{Accepted XXX. Received YYY; in original form ZZZ}

\pubyear{2026}

\begin{document}
\label{firstpage}
\pagerange{\pageref{firstpage}--\pageref{lastpage}}
\maketitle
\begin{abstract}
Higher-order statistics are becoming increasingly useful in clustering studies of the large-scale structure (LSS) of the universe. The $k$NN-CDFs are one such higher-order statistic, sensitive to all N-point correlations and easy to compute on data. Utilising their statistical gains for cosmological inference requires modelling them accurately for the tracer of interest. We use the hybrid Lagrangian bias expansion to enable this modelling in redshift space, building and validating the pipeline on \textit{Quijote} simulations. We show that the $k$NN-CDFs of a luminous red galaxy (LRG)-like sample can be modelled down to scales as small as $r=10\,\mathrm{Mpc}/h$, with the minimum reliable scale increasing with nearest-neighbour order. Moreover, the model parameters inferred from the $k$NN-CDFs also describe the galaxy power spectrum $P(k)$, indicating model consistency across the two statistics and the feasibility of combining them.
Fisher forecasts for galaxy auto clustering show that combining $P(k)$ and $k$NN-CDFs improves constraints on all parameters relative to either statistic alone. Relative to $P(k)$, the constraints on $b_3$ improve the most, while those on $\sigma_8$ and $\Omega_m$ improve by factors of 3.5 and 1.7, respectively. Including galaxy--matter cross clustering alongside galaxy auto clustering further improves the Fisher constraints from both $P(k)$ and $k$NN-CDFs, with their combination yielding the tightest constraints overall. Relative to $P(k)$, the largest improvement from this combination is found for $b_2$, while the constraints on $\sigma_8$ and $\Omega_m$ improve by factors of 4.2 and 1.7, respectively.
These results demonstrate the applicability of the hybrid Lagrangian bias framework to model the $k$NN-CDFs of biased tracers in redshift space, and the potential of combining higher-order and two-point statistics to extract cosmological information.
\end{abstract}




\section{INTRODUCTION}
\label{sec:intro}

The large-scale structure (LSS) of our universe and the underlying matter distribution contain crucial information about the cosmological model. Galaxy surveys such as SDSS \citep{SDSS1}, DESI \citep{DesiCollab2016}, and Euclid \citep{Euclid} have observed the galaxy distribution over large sky areas and continue to do so, enabling precise clustering measurements across a wide range of scales. A newer generation of facilities, such as the Vera C. Rubin Observatory's Legacy Survey of Space and Time \citep{LSST}, the Nancy Grace Roman Space Telescope \citep{Roman1,Roman2}, and SPHEREx \citep{SPHEREX}, will extend these observations to provide larger datasets with greater precision. In parallel, advances in high-resolution $N$-body simulations have led to a robust understanding of the evolution of the underlying dark matter field, the formation of halos, and their clustering (see e.g. \cite{Gadget2,Nbody1,Abacus1,Quijote2020}). Together, these observational and computational developments constitute some of the most powerful tools in the era of precision cosmology.

We observe the LSS through luminous, biased tracers of matter such as galaxies, quasars, and line emission from neutral hydrogen, among others. However, the dominant component shaping the LSS is dark matter, which, unlike galaxies, cannot be directly observed or imaged \citep{LSS_ref,Springel_LSS}. Inferring the matter distribution and thereby constraining cosmological parameters from observations requires a framework that relates galaxies (or any luminous tracer) to the underlying dark matter field (see e.g. \cite{Wechsler2018}). 
In principle, the connection between galaxies and the matter field can be obtained from hydrodynamical simulations that model both dark matter and baryonic physics. At present, some of the best hydrodynamical simulations include EAGLE \citep{EAGLE}, IllustrisTNG \citep{IllTNG1}, FLAMINGO \citep{Flamingo}, and SIMBA \citep{SIMBA}. However, no single simulation suite can reproduce all observed galaxy properties, and all of these simulations adopt slightly different prescriptions for subgrid physics (see e.g. \cite{Schaye2015,Dave2019,Diemer2019,Stevens2019}). Moreover, running such simulations for large volumes and across a wide range of cosmologies is computationally expensive.

As a result, semi-analytical models (see e.g. \cite{SAM1,SAM2,SAM3,SAM4,SAM5,SAM6,SAM7}) and empirical approaches such as Subhalo Abundance Matching (SHAM; see e.g. \cite{SHAM1,SHAM2,SHAM3,SHAM4,SHAM5}) and Halo Occupation Distribution (HOD; see e.g. \cite{HOD1,HOD2,HOD3,HOD4,HOD5,HOD6}) are commonly used to model the tracer--matter connection. Although these approaches are well developed and widely used, modelling observed galaxy clustering across cosmologies and model parameter values remains computationally expensive. For example, HOD-based modelling involves populating dark matter halos with galaxies according to a set of HOD parameters. Exploring the parameter space therefore requires generating galaxy catalogues for different HOD parameter sets and cosmologies.

A more generic and theoretically well motivated description of the tracer--matter connection is provided by the perturbative bias expansion (see \cite{LPT1} for a detailed review). In this framework, the tracer density field is expressed as a perturbative expansion in large-scale operators constructed from the linear matter field, truncated at a given order. This enables an accurate description of tracer clustering on linear to quasi-linear scales. The bias expansion can be formulated either in Eulerian space, corresponding to the spatial coordinates at the final redshift of interest (see e.g. \cite{Eulbias1,Eulbias2,Eulbias3,DAmico2020}), or in Lagrangian space, corresponding to coordinates at some initial redshift (see e.g. \cite{LPT1,LPT2,LPT3,Matsubara2008,Vlah2016,EulLPTbias}).
We adopt the latter formulation, where the biasing scheme is defined in Lagrangian coordinates as an expansion in a basis of symmetry-allowed operators, truncated at a given order in perturbation theory. The coefficients of this expansion, known as bias parameters, capture the response of a particular set of tracers to the corresponding basis operator field. As a result, different tracer populations correspond to different values of these bias parameters, which are typically marginalised over when inferring cosmological parameters. To make predictions in Eulerian coordinates, where the observables are measured, the initial Lagrangian biasing must be mapped to Eulerian space through the displacement field. Using $N$-body simulations, rather than perturbative approximations, to perform this mapping defines the hybrid Lagrangian bias framework. This framework therefore combines a perturbative description of biasing along with gravitational evolution obtained from $N$-body simulations. This approach has been shown to accurately model halo and galaxy power spectra up to $k = 0.6\, h/\mathrm{Mpc}$ \citep{Modi2020_SimAndSymt,Kokron2021a,Zennaro2022,Zennaro2023}.

The framework is therefore well suited to modelling the clustering of biased tracers such as galaxies. However, in spectroscopic surveys, galaxy positions are inferred from observed redshifts, which are affected by both cosmic expansion and the peculiar velocities of galaxies along the line of sight (LoS). This gives rise to redshift space distortions (RSDs), which make the observed galaxy distribution anisotropic \citep{RSDs1,Rsd3,Rsd2,RsdModel_old}. Models that connect observables to the underlying matter field must therefore account for RSDs. These distortions are not merely a nuisance to remove, but also provide a way to measure the growth of structure and offer a powerful test of gravity, a prospect made increasingly relevant by the precision of modern galaxy surveys (see e.g. \cite{Hawkins2003,Guzzo2008,Blake2011,Beutler2012,Samushia2012,Taruya2014,koyama2016cosmological,ishak2019testing,Icaza2020,Tamone2020,alam2021towards}).
To account for these effects, the hybrid Lagrangian bias framework is supplemented with a model for RSDs. This is done by incorporating peculiar velocities into the real-space tracer field predicted by the framework, accounting for both large-scale coherent motions that give rise to the Kaiser effect and small-scale intra-halo velocities that lead to the Finger-of-God (FoG) effect (see e.g. \cite{Jackson1972,Sargent1977,Davis1983,Guzzo1997,Hamilton1998,Orsi2018}). This extension has been shown to reproduce the galaxy $P(k)$ in redshift space on scales comparable to those reached in real space \citep{Pellejero2022,Pellejero2023}.

An important consideration, in addition to accurately modelling the tracer field, is the choice of statistic used to extract information from the galaxy distribution. Studies of LSS clustering have traditionally relied on two-point statistics, namely the correlation function $\xi(r)$ and the power spectrum $P(k)$, which fully characterise a Gaussian random field. The matter field is expected to be close to Gaussian at high redshifts and even at low redshifts on sufficiently large scales. However, on smaller scales, gravitational evolution drives the matter distribution away from Gaussianity. In these regimes, two-point statistics miss a significant amount of information contained in higher-order correlations, and thus do not fully exploit the constraining power of current survey datasets. Extracting this information requires going beyond two-point statistics, motivating numerous studies aimed at developing and utilising higher-order statistics (see e.g. \cite{Feldman2001,Verda2002,Gaztanaga2009,GilMarin2017,Philcox2022,Jiamin2024,Paillas2023,Uhlemann2020,Banerjee2021a,Hahn2021,Regaldo2024,Valogiannis2024}).

The higher-order statistics of interest in this work are the $k$-nearest-neighbour cumulative distribution functions ($k$NN-CDFs), which were first introduced in \cite{Banerjee2021a}, where their ability to capture auto clustering information beyond that contained in two-point statistics was demonstrated.
The $k$NN-CDFs in their most basic form are concerned with the CDF of distances to the $k^{\rm th}$ nearest neighbour data point (e.g. halos or galaxies). These are sensitive to all $N$-point correlations present in a dataset, and their measurement is computationally inexpensive. The formalism has since been extended to study cross-correlations between discrete tracers \citep{Banerjee2021b}, as well as between a discrete tracer population and a continuous field \citep{Banerjee2023}.

In recent studies, the $k$NN-CDF formalism has been applied to simulations in a variety of contexts, including neutral hydrogen--galaxy cross-correlations in the post-reionisation regime \citep{Chand2025} and also during the epoch of reionisation (EoR; \citealt{Chakraborty2026}), and to infer cosmological parameters from halos using neural networks \citep{Chatterjee2026}. Observational applications include auto clustering of redMaPPer clusters \citep{Wang2022}, the clustering of gravitational-wave sources with the LSS \citep{Gupta2024}, cross-correlations between IceCube neutrinos and astrophysical sources \citep{Zhou2025}, and the CDFs of weak-lensing convergence fields from DES Y3 \citep{Anbajagane2023}. 
More recently, \cite{Gangopadhyay2025} provided a geometric interpretation of $k$NN-CDFs, showing that they capture nearly the same information as Minkowski functionals while being significantly faster to compute.
The first study to model the $k$NN-CDFs of biased tracers was presented in \cite{Banerjee2022}, where the authors used the hybrid Lagrangian bias framework to model the real-space clustering of dark matter halos. Extending such modelling to redshift space is important for applying the $k$NN-CDFs to spectroscopic galaxy samples. In a related context, \cite{Yuan2024} modelled the 2D $k$NN-CDFs of galaxies in redshift space using an HOD-based forward model and an emulator to constrain cosmological parameters.

In this work, we develop a pipeline based on the hybrid Lagrangian bias expansion to model the $k$NN-CDFs of biased tracers in redshift space and validate it on dark matter halos and galaxies in simulations. We also test the consistency of the model across the $k$NN-CDFs and $P(k)$, and perform Fisher analysis to quantify the gains in constraining model and cosmological parameters from combining the two statistics.
The rest of the paper is organised as follows. In Section~\ref{sec:tracer_matter}, we describe the theoretical framework of the hybrid Lagrangian bias expansion and the model for RSDs. In Section~\ref{sec:knnmodel}, we describe the computation of the $k$NN-CDFs, including their measurement from data and the corresponding model predictions. Section~\ref{sec:implement} details the simulations and galaxy catalogue used in the modelling analysis and describes the pipeline's implementation. In Section~\ref{sec:ModellingResults}, we present the modelling results for the galaxy $k$NN-CDFs and $P(k)$. In Section~\ref{sec:Fisher}, we present the Fisher analysis of the $k$NN-CDFs, $P(k)$, and their combination. Finally, in Section~\ref{sec:Conclusions}, we summarise our main findings and discuss the future directions.

\section{Tracer--matter connection}\label{sec:tracer_matter}
\subsection{Hybrid Lagrangian Bias Expansion}\label{sec:HEFTmodel}
The hybrid Lagrangian bias expansion approach to model the tracer field at late times involves combining the bias expansion defined in Lagrangian space with the displacement field from $N$-body simulations. This hybrid method extends the reach of the bias expansion to model the clustering statistics of tracers down to smaller, mildly non-linear scales.

In this framework, the proto-tracer overdensity $\delta_T(\vct{q})$, where $\vct{q}$ denotes the Lagrangian coordinates, is expressed as \citep{Vlah2016,Matsubara2008,Modi2020_SimAndSymt,Kokron2021a,Zennaro2022}
\eq{LagrFunc}{
\delta_T(\vct{q}) = F\big[\delta_L(\vct{q}),\, s_{ij}(\vct{q})\big] + \epsilon(\vct{q}) \, ,
}
where $\delta_L(\vct{q})$ denotes the linear matter overdensity field, and $s_{ij}(\vct{q}) = (\partial_i \partial_j - \frac{1}{3}\delta^{K}_{ij}\,\nabla^2) \phi(\vct{q})$ denotes the traceless tidal tensor ($\phi$ being the gravitational potential). The $\epsilon(\vct{q})$ term represents a stochastic component that accounts for the fact that the bias expansion might not fully determine the tracer field.

The functional $F$ is expanded in the basis of large-scale operators $O_L$ consistent with symmetries of the underlying physical system. Up to quadratic order, the relevant operators include the linear matter overdensity $\delta_L(\vct{q})$, its square $\delta_L^2(\vct{q})$, the shear operator $s_L^2(\vct{q}) = s_{ij}s^{ij}$, and the Laplacian operator $\nabla^2 \delta_L(\vct{q})$. 
Including higher-order operators can improve the recovery of leading-order bias parameters and also help isolate stochastic contributions \citep{Shiferaw2025}. However, as the number of parameters increases, the model complexity also increases. In previous studies of modelling the $P(k)$ of halos and galaxies, quadratic bias expansion is found to be sufficient (see e.g. \cite{Vlah2016,Modi2020_SimAndSymt,Zennaro2022}). In our modelling of the $k$NN-CDFs of biased tracers, we find it adequate to include the next-order operator $\delta_L^3$ in the expansion. At cubic order, the bias expansion includes a few additional operators (see e.g. \cite{Saito2014,LazeyrasSchmidt2018}), which are known to contribute degenerately to the tracer power spectrum at one-loop order. While this degeneracy can, in principle, be lifted using higher-order statistics \citep{LazeyrasSchmidt2018}, a complete exploration of the full cubic operator basis for modelling the $k$NN-CDFs is beyond the scope of this study and will be taken up in future work. The resulting bias expansion takes the form
\eq{BiasExpansion}{
F[\delta_L(\vct{q}), s_{ij}(\vct{q})] &\approx
 1 + b_1 \delta_L(\vct{q})
+ b_2\big(\delta_L^2(\vct{q}) - \langle \delta_L^2 \rangle\big)\\
&\quad+ b_{s^2}\big(s_L^2(\vct{q}) - \langle s_L^2 \rangle\big) 
+ b_{\nabla^2} \nabla^2 \delta_L(\vct{q}) \\ 
&\quad+ b_3 \,\big( \delta_L^3(\vct{q}) - 3\sigma^2 \delta_L(\vct{q}) \big)\, .
}
The stochastic field $\epsilon(\vct{q})$ is assumed to satisfy $\langle \epsilon(\vct{q}) \rangle = 0$ and $\langle \epsilon(\vct{q})\, O_L(\vct{q}) \rangle = 0$.

To predict the tracer overdensity at late times (low redshift), we need to map the initial and final coordinates of tracers. This mapping is given by
\eq{displacement}{
\vct{x}(\vct{q},z) &= \vct{q} + \vct{\Psi}(\vct{q},z) \, ,
}
where $\vct{\Psi}(\vct{q},z)$ is the displacement field, a function of the initial coordinates and the final redshift. The key aspect of the hybrid framework, as mentioned earlier, is that the displacement field is taken from $N$-body simulations rather than using perturbative approximations.
By mass conservation, the late-time matter overdensity can be written as
\eq{deltam_x}{
1 + \delta_m(\vct{x},z)
&= \int d^3 \vct{q} \;
\delta_{\rm D}\big(\vct{x} - \vct{q} - \vct{\Psi}(\vct{q},z)\big) \, ,
}
where $\delta_{\rm D}$ denotes the Dirac delta function. Following this prescription, the late-time tracer field is obtained by advecting the initial tracer weights, defined by the Lagrangian bias expansion (equation~\ref{BiasExpansion}), to the redshift of interest under the mapping given by $\vct{\Psi}$
\eq{Eul_ovden}{
1 + \delta_T(\vct{x},z) = \int d^3 \vct{q} 
\left[ F[\delta_L(\vct{q}), s_{ij}(\vct{q})] + \epsilon(\vct{q}) \right] \,\times \\
\quad \delta_{\rm D}\big(\vct{x} - \vct{q} - \vct{\Psi}(\vct{q},z)\big) \, .
}
The operator fields can also be advected separately and then combined with the bias coefficients to construct $\delta_T(\vct{x},z)$.

\subsection{Redshift space distortions}\label{sec:Rsdmodel}
In spectroscopic observations, galaxies are mapped in redshift space, and their distribution is therefore affected by RSDs, which we need to account for in our modelling. Our treatment of RSDs follows the framework presented in \cite{Pellejero2022}. The distortions caused by large-scale coherent flows are accounted for by displacing the tracer coordinates at the final redshift along the LoS according to
\eq{Rsd1}{
\vct{s} = \vct{x} + \frac{\vct{v}_{\rm tr}(\vct{x})\!\cdot\!\hat{\vct{z}}}{aH}\,\hat{\vct{z}} \, ,
}
where $\vct{s}$ denotes the redshift-space coordinate, $\hat{\vct{z}}$ is the LoS direction, and $\vct{v}_{\rm tr}(\vct{x})$ is the tracer velocity. For tracers within a halo, we use the velocity of the host dark matter halo as the tracer velocity, while for tracers outside a halo, we use the particle's own velocity. This reflects the fact that tracers residing in halos sample the underlying dark matter phase-space distribution differently from individual dark matter particles. Their velocities are therefore more appropriately represented by the bulk velocity of their host halo, whereas particles outside halos trace the local velocity field.
This step modifies the displacement field as
\eq{RsdPhi}{\vct{\Psi}^{\,s} = \vct{\Psi} + \frac{\vct{v}_{\rm tr}(\vct{x})\!\cdot\!\hat{\vct{z}}}{aH}\,\hat{\vct{z}} \, ,
}
where the explicit dependence on $\vct{q}$ has been suppressed for notational brevity.
This procedure is called central distortion, where all tracer points within a halo are mapped using the velocity of that halo. We denote the tracer overdensity after this step by $\delta_{T}^{\rm cd}(\vct{s})$. Since this procedure removes intra-halo velocities, the resulting description is suitable for samples consisting only of host halos or central galaxies.

In addition to this coherent motion, galaxies within virialised structures exhibit random orbital motions with a range of velocities characterised by velocity dispersion, giving rise to the Finger-of-God (FoG) effect. This phenomenon leads to elongation of the galaxy distribution along the LoS and a suppression of clustering on small scales. We model this effect by convolving the centrally distorted field with an exponential kernel describing the relative LoS displacements of satellite galaxies. The kernel takes the form
\eq{PDFvrel}{
K_{\rm FoG}(s_z) = \frac{\lambda_{\rm FoG}}{2}
\exp\!\left(-\lambda_{\rm FoG}|s_z|\right)\,,
}
where $\lambda_{\rm FoG}$ controls the damping strength, with larger values corresponding to weaker damping and vice versa \citep{Orsi2018}. Denoting the satellite fraction by $f_{\rm sat}$, such that the central fraction is $(1-f_{\rm sat})$, the resulting tracer overdensity field in redshift space can be written as
\eq{FoG_eff}{
\delta_{T}^{\rm FoG}(\vct{s}) =
(1-f_{\rm sat})\,\delta_{T}^{\rm cd}(\vct{s})
+ f_{\rm sat} \left[\delta_{T}^{\rm cd}(\vct{s}) *_z
K_{\rm FoG}(s_z)\right] \, ,
}
where $*_z$ denotes convolution along the $z$-axis (LoS direction). With the correct treatment of RSDs, the same set of bias parameters should be able to model both real- and redshift-space measurements, since RSDs alter the observed clustering but do not change the underlying tracer bias.

\section{Nearest Neighbour Distributions}\label{sec:knnmodel}
The $k$NN-CDFs are sensitive to the spatial clustering information at all orders, rather than being limited to second-order (power spectrum) or third-order (bispectrum) correlations, while remaining computationally efficient to compute \citep{Banerjee2021a}. This makes them particularly useful for studying the clustering of LSS tracers on small scales, where the matter field is highly non-Gaussian. 
The $k$NN-CDF formalism has been developed for measuring the auto clustering of discrete tracers, cross correlations between two populations of discrete tracers, and cross correlations between discrete tracers and a continuous tracer field. Below, we briefly describe the methodology for measuring the auto clustering of discrete tracers and their cross clustering with a continuous field, and present the corresponding $k$NN-CDF expressions for obtaining their predictions.

\subsection{Measurements}\label{sec:kNNMeas}
The $k$NN-CDF measurement for the auto clustering of a set of $N_d$ discrete tracers, and their cross clustering with a continuous tracer field (e.g. the matter or neutral hydrogen field) with overdensity $\delta(\vct{x})$, both residing within a cosmological box, is done as follows:
\begin{itemize}
    \item We first sample the given volume with a large number $N_q$ of randomly positioned query points, ensuring $N_q \gg N_d$.

    \item The CDF of distances from these query points to their $k^{\rm th}$ nearest discrete tracer (data point) defines $\mathrm{CDF}_{k\mathrm{NN}}(r)$, which also corresponds to the probability of finding at least $k$ tracers within a distance $r$ of a query point, denoted by $P_{\geq k}(r)$.
    
    \item This definition naturally extends to the cross clustering between discrete tracers and a continuous field, giving rise to the joint $k$NN-CDFs, which involve computing the joint probability of (i) at least $k$ tracers lying within a distance $r$ of a query point; and (ii) the smoothed continuous field (using a top-hat kernel of radius $r$), interpolated at the same query point, satisfying a certain threshold criterion. We denote this joint probability by $\mathrm{JCDF}_{k\mathrm{NN}}(r)$, and this statistic is sensitive to the cross correlations between the tracers and the field at all orders \citep{Banerjee2023}.
    
    \item For the joint $k$NN-CDF analysis, we choose to impose a percentile-based threshold $\delta_r^*$ on the smoothed field, such that the selection criterion is $\delta_r < \delta_r^*$.
\end{itemize}

\subsection{Predictions}\label{sec:kNNPred}
Consider a set of discrete tracers with mean number density $\bar{n}$, representing a local Poisson sampling of an underlying density field $\rho_T = (1+\delta_T)\,\bar{\rho}_T$, with local rate $\lambda(\vct{x}) = \bar{n}\,[1+\delta_T(\vct{x})]$. The probability of finding exactly $k$ of these tracers within a spherical volume $V$ of radius $r$ centred at position $\vct{x}$ is given by \citep{Banerjee2022}

\eq{Prob1}{
\mathcal{P}_{k}(r)\big|_{\vct{x}} = 
\frac{\big[\bar{n}V \big(1+\delta_{T,r}^{\,\vct{x}} \big)\big]^k 
\exp\!\big[-\bar{n}V \big(1+\delta_{T,r}^{\,\vct{x}} \big)\big]}{k!} \, ,
}

where $\delta_{T,r}^{\,\vct{x}}$ denotes the tracer overdensity at position $\vct{x}$ (corresponding to a grid point in the simulation box), smoothed on scale $r$. The above expression follows directly from the Poisson probability formula. 
The volume-averaged probability then becomes
\eq{Prob2}{
\mathcal{P}_{k}(r) = \left\langle \mathcal{P}_{k}(r)\big|_{\vct{x}} \right\rangle \, .
}

Following the above expressions, the CDF of distances to the $k^{\rm th}$ nearest neighbour is given by
\eq{CDF_pred}{
\mathrm{CDF}_{k\mathrm{NN}}(r) &= \!\mathcal{P}_{\geq k}(r) 
=\! 1 - \sum_{m=0}^{k-1} \! \mathcal{P}_m(r) \\
&= \left\langle \! 1 -  \sum_{m=0}^{k-1} 
\frac{\big[\bar{n}V \big(1+\delta_{T,r}^{\,\vct{x}} \big)\big]^m
\exp\!\big[-\bar{n}V \big(1+\delta_{T,r}^{\,\vct{x}} \big)\big]}{m!}\!
\right\rangle \, .
}

The cross clustering between the discrete tracers and a field denoted by $\delta$, within the $k$NN-CDF formalism, is given by the joint probability of finding at least $k$ tracers within a sphere of radius $r$ (centered at $\vct{x}$) and the smoothed field $\delta_r^{\vct{x}}$, obtained using a top-hat kernel of radius $r$, satisfying the specified criterion. Consistent with the choice made for the measurements, the criterion is percentile-based and is implemented through the Heaviside function $\Theta(\delta_r^* - \delta_r^{\vct{x}})$. The expression for the joint $k$NN-CDF is thus given by 
\eq{JCDF_pred}{
\mathrm{JCDF}_{k\mathrm{NN}}(r) & = \bigg \langle \! \Big( 1 \!- \! \sum_{m=0}^{k-1} \!
\frac{\big[\bar{n}V \big(1+\delta_{T,r}^{\,\vct{x}} \big)\big]^m
\exp\!\big[-\bar{n}V \big(1+\delta_{T,r}^{\,\vct{x}} \big)\big]}{m!} \Big) \\ 
&\quad \times \Theta( \delta_r^* - \delta_{r}^{\vct{x}}) \!\bigg \rangle \, .
}

\section{Implementation details}\label{sec:implement}
\subsection{Simulation and tracer catalogues}\label{sec:sim_tracers} 
We utilise the \textit{Quijote} simulations \citep{Quijote2020} to construct our modelling pipeline and validate it on dark matter halos and galaxies from the same suite at a redshift of $z=0.5$, relevant to the low-redshift regime probed by current spectroscopic surveys. We work with the high-resolution fiducial cosmology suite of these simulations, which has a box length of $L=1\,\mathrm{Gpc}/h$ and contains $1024^3$ cold dark matter (CDM) particles. This fiducial cosmology adopts $\Omega_m=0.3175$, $\Omega_b=0.049$, $h=0.6711$, $n_s=0.9624$, $\sigma_8=0.834$, and $M_{\nu}=0$. We focus on explaining the galaxy catalogues here and present the corresponding modelling results in Section~\ref{sec:ModellingResults}. The halo catalogue and its modelling results are explained in Appendix~\ref{sec:HaloResults}.

We generate the halo catalogues using the \textsc{Rockstar} halo finder \citep{Behroozi2013} and the galaxies are obtained using the HOD prescription. The HOD framework relies on a probabilistic mapping between halo mass and the number of central and satellite galaxies hosted by that halo. We implement the HOD using the \texttt{Halotools} package \citep{Hearin2017}\footnote{\url{https://halotools.readthedocs.io/en/latest/}}. The central galaxy occupation number follows a Bernoulli distribution, while the satellite occupation number follows a Poisson distribution. The corresponding mean occupation functions are given by (see e.g. \cite{Zheng2005,Zheng2007})

\eq{HOD_Cen}{
\langle N_{\rm cen}(M) \rangle =
\frac{1}{2}\left[1 + \mathrm{erf}\!\left(
\frac{\log M-\log M_{\rm min}}{\sigma_{\log M}}
\right)\right],
}
\eq{HOD_Sat}{
\langle N_{\rm sat}(M) \rangle =
\left(\frac{M-M_0}{M_1}\right)^{\alpha},
}

where $\langle N_{\rm cen}(M)\rangle$ and $\langle N_{\rm sat}(M)\rangle$ denote the mean numbers of central and satellite galaxies, respectively, hosted by a halo of mass $M$.
The parameter $M_{\rm min}$ sets the characteristic halo mass for central galaxy occupation, while $\sigma_{\log M}$ controls the width of this transition. The parameter $M_0$ defines the cut-off mass above which halos can host satellite galaxies, $M_1$ sets the characteristic halo mass scale for hosting satellite galaxies, with the mean satellite occupation reaching one at $M=M_0+M_1$, and $\alpha$ determines the slope of the satellite occupation as a function of halo mass.

\begin{figure}
    \centering
    \includegraphics[width=0.7\linewidth]{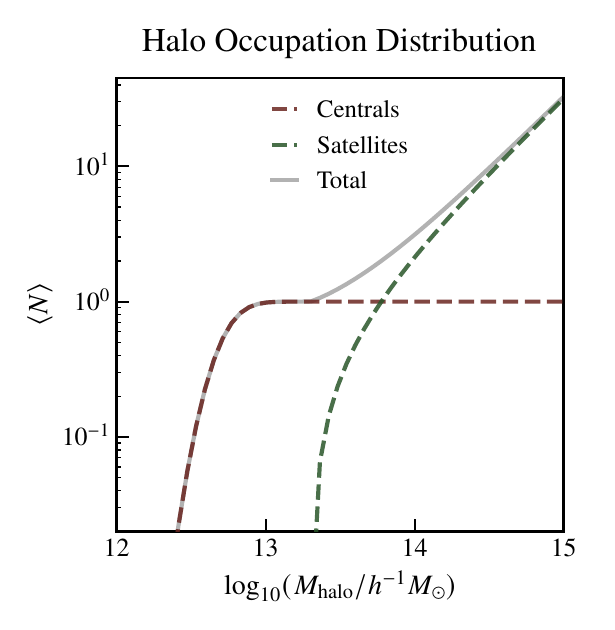}
    \caption{The halo occupation distribution (HOD) of the LRG-like galaxies constructed using \textit{Quijote} simulations at a redshift of $z=0.5$, and used for our main modelling analysis (presented in Section~\ref{sec:ModellingResults}). The maroon dashed curve here denotes the central galaxy occupation, while the green dashed curve denotes the satellite galaxy occupation, and the total occupation is shown by the grey solid curve; these occupations are shown as a function of host halo mass. The HOD parameter values are $\log M_{\rm min}=12.7$, $\sigma_{\log M}=0.2$, $\log M_0=13.3$, $\log M_1=13.6$, and $\alpha=1.08$. See Section~\ref{sec:sim_tracers} for details of the HOD prescription.}
    \label{fig:GalxHOD_1024p}
\end{figure}

We set $\log M_{\rm min}=12.7$, $\sigma_{\log M}=0.2$, $\log M_0=13.3$, $\log M_1=13.6$, and $\alpha=1.08$, chosen to yield a luminous red galaxy (LRG)-like population (the parameter values are close to those reported by \cite{Yuan2024_LRG}). Figure~\ref{fig:GalxHOD_1024p} depicts the HOD for this galaxy catalogue, and the number density of the galaxy sample that we work with is $\bar{n}_g=9.52\times10^{-4}\,(\mathrm{Mpc}/h)^{-3}$.

\subsection{Analysis pipeline}\label{sec:pipeline}
The general procedure for measuring the auto and cross clustering within the $k$NN-CDF formalism is described in Section~\ref{sec:kNNMeas}, and we use $N_q=256^3$ query points for the measurements. To compute the galaxy--matter cross clustering via joint $k$NN-CDFs, the matter overdensity field is defined on an $N_g=1024^3$ grid, and we choose the $90^{\mathrm{th}}$ percentile value of the smoothed matter field to define $\delta_r^*$. We average over 10 simulation realisations to assess the agreement between data and model predictions.
The $k$NN-CDF signal for a given tracer sample depends directly on its number density (see equation~\ref{CDF_pred}), so for a sufficiently dense sample, the CDF can rapidly approach unity at relatively small separations. In such cases, we work with subsamples of the tracer catalogue whose effective number density allows the signal to vary appreciably over the scales of interest.

In this work, we focus on measurements of the $k$NN-CDFs over scales ranging from $10\,\mathrm{Mpc}/h$ to $30\,\mathrm{Mpc}/h$, which span the quasi-linear regime, with increasingly nonlinear behaviour towards smaller scales. We work with eight subsamples of the galaxy catalogue, leading to an effective number density of $\bar{n}_g^{\rm eff} = 1.19\times10^{-4}\,(h/\mathrm{Mpc})^{3}$, and use the mean of the $k$NN-CDF measurements across these subsamples. In contrast, for the power spectrum, the tracer number density affects only the noise and does not modify the shape of the clustering contribution to $P(k)$; downsampling is therefore not required.
The two statistics also differ in the clustering information they access. The $k$NN-CDFs are sensitive to all $N$-point correlations and are therefore particularly informative beyond two-point statistics on quasi-linear to nonlinear scales, where the clustering is non-Gaussian. Our analysis of $P(k)$, however, extends to much larger scales ($k\approx 0.015 \, h/\mathrm{Mpc}$) than those used for the $k$NN-CDFs, and these scales carry additional information relevant for some of the cosmological parameters.
A further difference concerns the anisotropic clustering imprinted by RSDs. The $k$NN-CDFs as formulated here are angle-averaged: they do not capture the LoS dependence of the clustering and respond only to the net change in the overall clustering, whereas the $\ell>0$ multipoles of $P(k)$ capture the anisotropic part directly. We therefore expect the $k$NN-CDFs to be less sensitive to the FoG parameters. Extending the $k$NN-CDF formalism to capture the angular dependence of the clustering is beyond the scope of this work and is deferred to future work.

The detailed implementation of the hybrid Lagrangian bias framework in our modelling pipeline is as follows.

The operator fields entering the bias expansion, namely $\delta_L$, $\delta_L^2$, $s_L^2$, $\nabla^2 \delta_L$, and $\delta_L^3$, are constructed at an initial redshift of $z_i = 127$ on a grid with $N_g = 1024^3$ cells. To account for small-scale exclusion effects, the operator fields are smoothed with a Gaussian filter. We adopt a smoothing scale of $k_s = 0.75\,\ihmpc$, chosen to be comparable to the typical Lagrangian size of dark matter halos \citep{Zennaro2022}. Since the $k$NN-CDFs are configuration space statistics, we implement the smoothing in configuration space using a Gaussian kernel with smoothing scale $R = 1/k_s \!\approx\! 1.33\,\hmpc$, and for the $P(k)$ modelling we apply the equivalent Gaussian kernel in Fourier space.
The smoothed operator fields are evaluated at the initial particle positions, which sample the Lagrangian space. The value of each operator is rescaled using the linear growth factor $D(z)$ associated with the final redshift. An operator proportional to the $n^{\rm th}$ power of $\delta_L$ scales as
\[
g_n \equiv \left(\frac{D(z_f)}{D(z_i)}\right)^n.
\]
In particular,
\[
\delta_L \!\rightarrow\! g_1\delta_L,\;
\delta_L^2 \!\rightarrow\! g_2\delta_L^2,\;
s_L^2 \!\rightarrow\! g_2s_L^2,\;
\nabla^2\delta_L \!\rightarrow\! g_1\nabla^2\delta_L,\;
\delta_L^3 \!\rightarrow\! g_3\delta_L^3.
\]
Given the bias parameters, one can either combine the operators before advection or advect the individual components separately and combine them afterwards to obtain $\delta_T(\vct{x})$. In either case, the field is advected to the final redshift and deposited at the final particle positions, in real or redshift space depending on the scenario. The real-space construction requires only the particle positions and ID information at the final redshift, whereas the mapping to redshift space additionally requires the final particle velocities and the velocities of their host halos (as detailed in Section~\ref{sec:HEFTmodel}).

For $P(k)$ modelling, it is more efficient to compute the auto and cross power spectra of the individual operator fields and then combine them using the bias parameters $b_i$ to obtain $P(k)$ predictions. This follows from the linearity of the operations involved. The redshift-space $P(k)$ for a tracer population consisting of satellites can be written as a sum of auto and cross power spectra of the basis operators evaluated at the redshift-space coordinate $\vct{s}$ (equation~\ref{Rsd1}), multiplied by the FoG damping factor:
\eq{Rsd_Pk}{
\!\!
P_T(k,\mu)
=
\Bigg[(1-f_{\rm sat})
+f_{\rm sat}
\frac{1}{1+\left(\dfrac{k\mu}{\lambda_{\rm FoG}}\right)^2}\Bigg]^2\!
\times \!
\sum b_i b_j P^{\vct{\,s}}_{ij}
+P_{\epsilon},
}
where $\mu\equiv\cos\theta=k_\parallel/k$ is the cosine of the angle between $\vct{k}$ and the LoS direction, and $P_{\epsilon}$ denotes the stochastic contribution to the power spectrum, which at lowest order is modelled as ${c_1}/{\bar n}$ and contributes only to the monopole (see e.g. \cite{Perko2016, Zennaro2022}).
The inclusion of the $\delta_L^3$ operator along with the quadratic bias expansion gives a total of 21 auto and cross power spectra components. In real space, equation~\ref{Rsd_Pk} simplifies considerably: the FoG factor is absent, and the operator spectra $P_{ij}(k)$ are evaluated in real space, yielding a $\mu$-independent tracer power spectrum $P_T(k)$.

The $k$NN-CDF predictions, on the other hand, are obtained by combining the advected operator fields with the bias parameters to construct the late-time tracer overdensity field $\delta_{\rm T}$ in real space and $\delta_{\rm T}^{\rm cd}$ in redshift space (accounting for central distortion). The FoG effects are incorporated by modifying $\delta_{\rm T}^{\rm cd}$ using equation~\ref{FoG_eff}. For all cases considered, we find that the $k$NN-CDFs can be adequately modelled with $\epsilon(q)=0$, which means no stochastic contribution is required. Given the final tracer overdensity field, the $k$NN-CDF predictions can be computed using the expressions presented in Section~\ref{sec:kNNPred}.
The model parameters for the $k$NN-CDFs are inferred through an iterative fitting procedure, in which they are adjusted until satisfactory agreement with the measurements is achieved. A standard $\chi^2$ minimisation would require a faster prediction pipeline, for example via an emulator, which we intend to develop in future work.

To infer the bias parameters that best fit the tracer power spectrum, we minimise
\eq{eq:chisq}{
\chi^2 = \left(\mathbf{P}_d - \mathbf{P}_m\right)^{\mathrm{T}} \mathbf{C}^{-1} \left(\mathbf{P}_d - \mathbf{P}_m\right) \, ,
}
where $\mathbf{P}_d$ and $\mathbf{P}_m$ denote the data and model power spectrum vectors, respectively, and $\mathbf{C}$ denotes the covariance matrix. We assume a Gaussian covariance, retaining only the disconnected contribution (see e.g. \cite{Grieb2016,Li2019,Kokron2021a,Pellejero2022}). 

\begin{figure*}
    \centering
    \includegraphics[width=0.6\linewidth]{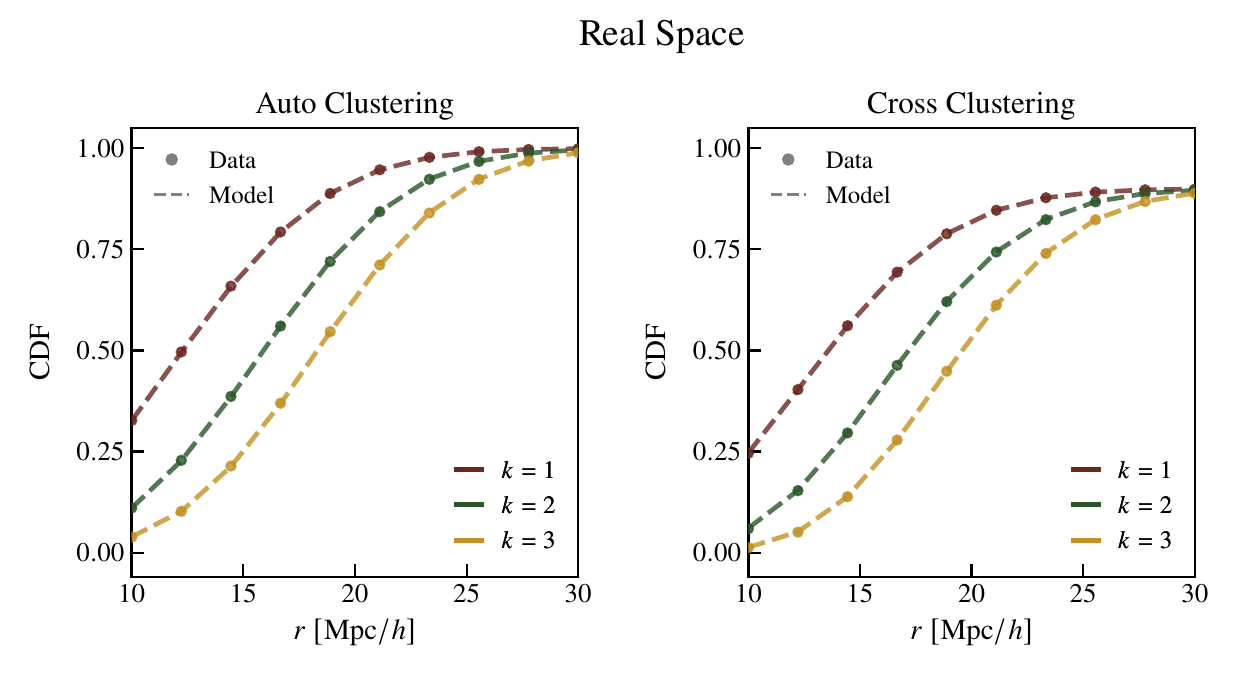}
    \caption{The real space $k$NN-CDFs for galaxy auto clustering (left panel) and galaxy--matter cross clustering (right panel). This sample is obtained using an HOD prescription for an LRG-like sample at a redshift of $z=0.5$ (see Section~\ref{sec:sim_tracers} for details). In each panel, the colour pattern follows: maroon for 1NN-CDF, green for 2NN-CDF, and golden yellow for 3NN-CDF. The circular points denote the $k$NN-CDF measurements on the galaxies, while the dashed lines show the model prediction using the hybrid Lagrangian bias framework, with both quantities averaged over 10 realisations. The model prediction closely follows the data across the scales shown. We present the corresponding residuals in Figure~\ref{fig:Real_LRG_knnCdf_1024p_Resid}.}
    \label{fig:Real_LRG_knnCdf_1024p}
\end{figure*}
\begin{figure*}
    \centering
    \includegraphics[width=0.85\linewidth]{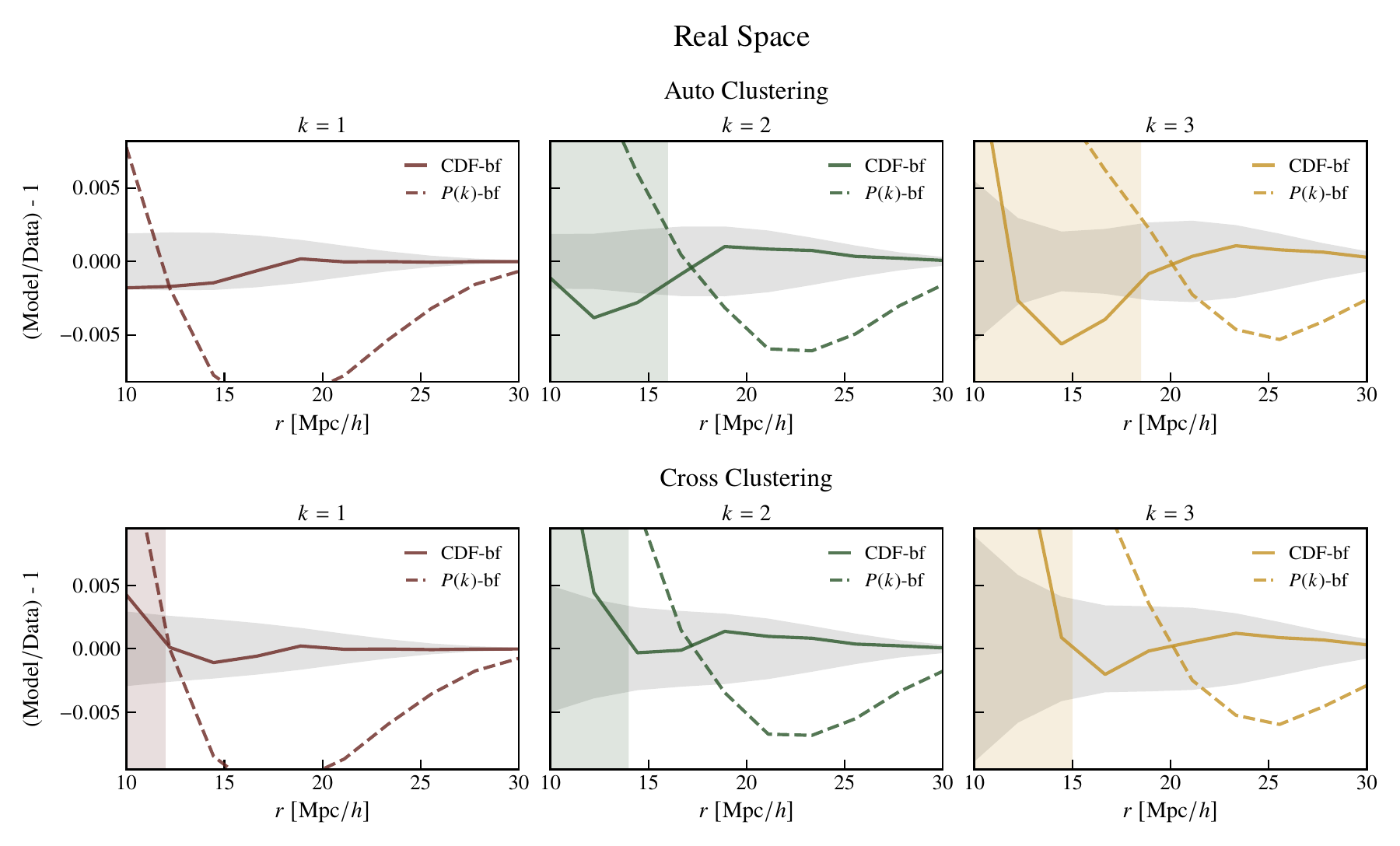}
    \caption{Residuals between the $k$NN-CDF measurements for galaxies in real space and the corresponding model predictions within the hybrid Lagrangian bias framework for the first three nearest neighbours ($k=1,2,3$). The top panel presents the galaxy auto clustering case, while the bottom panel presents the galaxy--matter cross clustering case. In each panel, the solid line shows the residuals for predictions obtained using the bias parameters inferred from the galaxy $k$NN-CDFs (corresponding to the dashed line in Figure~\ref{fig:Real_LRG_knnCdf_1024p}), while the dashed line shows the residuals for predictions obtained using the bias parameters inferred from the real-space galaxy auto and galaxy--matter cross power spectra. The grey shaded region represents the scatter in the $k$NN-CDF measurements across 100 realisations, whereas the shaded region extending up to a given $r$ in each panel marks the scales where the agreement between data and model is not good. The agreement extends down to $r=10\,\mathrm{Mpc}/h$ at best, for the 1NN-CDF in auto clustering.}
\label{fig:Real_LRG_knnCdf_1024p_Resid}
\end{figure*}
\section{Modelling Results}\label{sec:ModellingResults}
\begin{figure*}
    \centering
    \includegraphics[width=0.6\linewidth]{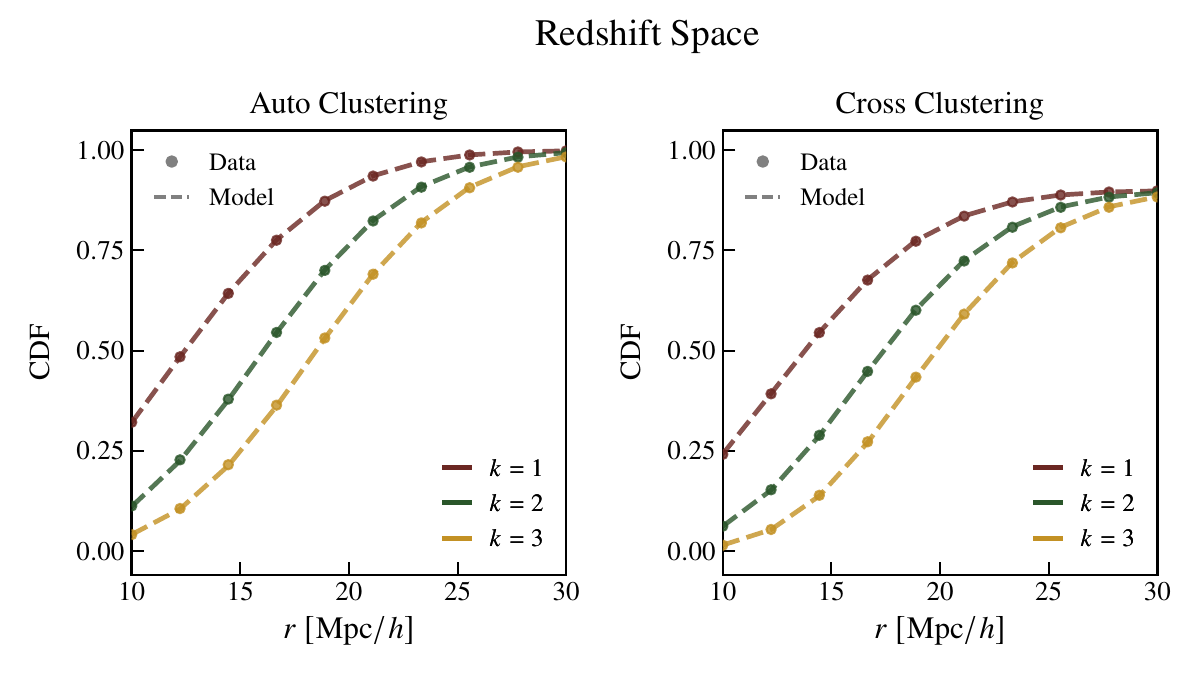}
    \caption{The $k$NN-CDFs for galaxy auto clustering (left panel) and galaxy--matter cross clustering (right panel), with galaxies in redshift space and the matter field in real space (refer to  Section~\ref{sec:ModellingResults} for details). In each panel, the colour pattern follows: maroon for 1NN-CDF, green for 2NN-CDF, and golden yellow for 3NN-CDF. The circular points denote the $k$NN-CDF measurements on the galaxies, while the dashed lines show the model prediction using the hybrid Lagrangian bias framework, with both quantities averaged over 10 realisations. The same set of bias parameters as in real space is used, with additional FoG parameters introduced through the RSD modelling. The model closely follows the data across the scales shown, indicating the consistency of both the bias expansion and the RSD model. The corresponding residuals are presented in Figure~\ref{fig:Rsd_LRG_knnCdf_1024p_Resid}.
    }
    \label{fig:Rsd_LRG_knnCdf_1024p}
\end{figure*}
\begin{figure*}
    \centering
    \includegraphics[width=0.85\linewidth]{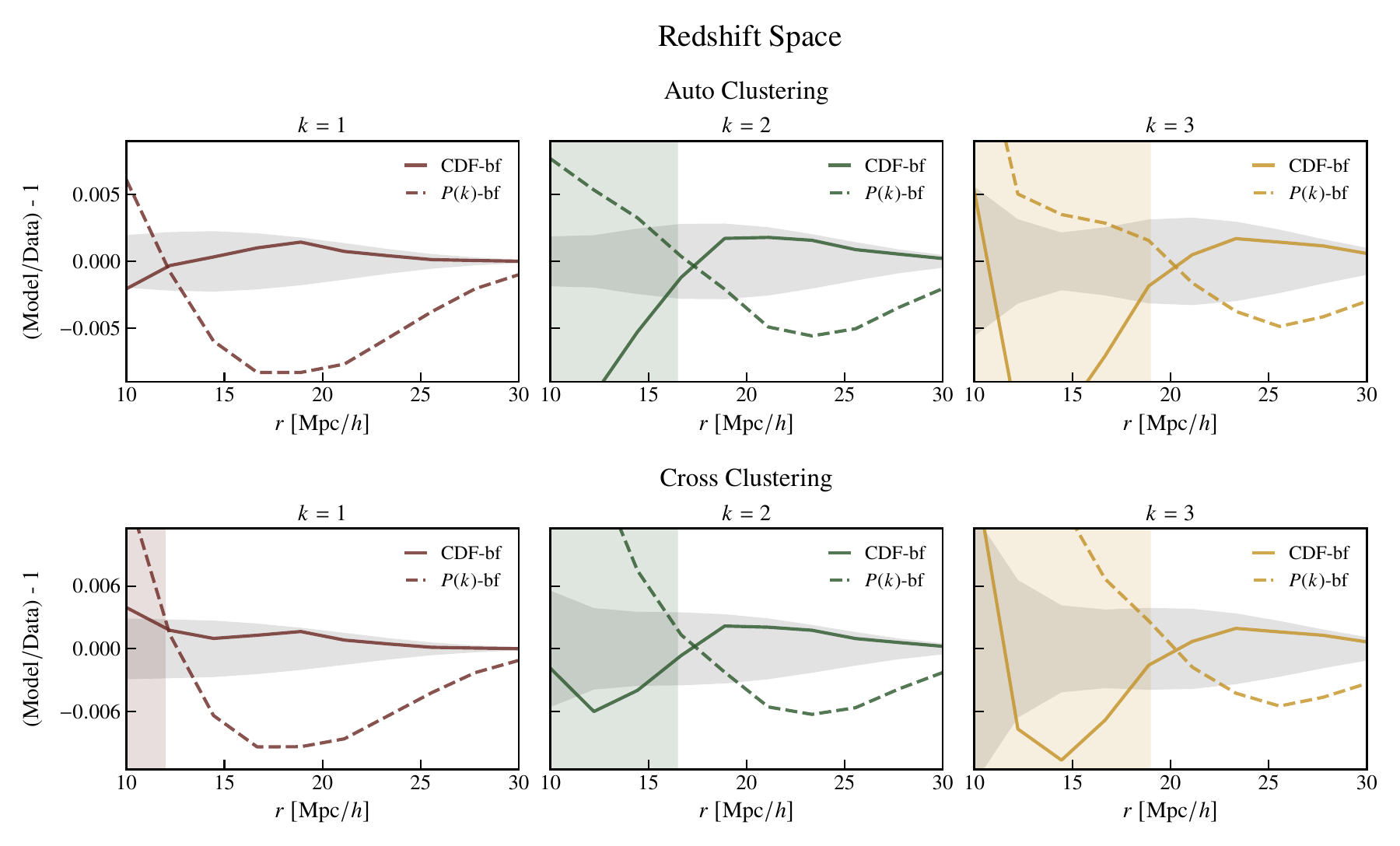}
    \caption{Residuals between the $k$NN-CDF measurements and model predictions for galaxies in redshift space, shown for the first three nearest neighbours ($k=1,2,3$). The top panel presents the galaxy auto clustering case, while the bottom panel presents the galaxy--matter cross clustering case (matter field in real space). In each panel, the solid line shows the residuals for predictions obtained using the bias parameters inferred from the real-space galaxy $k$NN-CDFs (same set corresponds to the solid line in Figure~\ref{fig:Real_LRG_knnCdf_1024p_Resid}), while the dashed line shows residuals for the predictions obtained using the bias parameters inferred from galaxy auto and galaxy--matter cross power spectra in real space (same set corresponds to the dashed line in Figure~\ref{fig:Real_LRG_knnCdf_1024p_Resid}). The grey shaded region represents the scatter in the measurements across 100 realisations, while the shaded region extending up to a given $r$ in each panel indicates the scales where the data and model do not agree well. The 1NN-CDF in auto clustering is again the best modelled, with agreement down to $r=10\,\mathrm{Mpc}/h$.}
    \label{fig:Rsd_LRG_knnCdf_1024p_Resid}
\end{figure*}

In this section, we present the results of our analysis aimed at modelling the $k$NN-CDFs of galaxies (LRG-like sample defined in Section~\ref{sec:sim_tracers}) within the framework of the hybrid Lagrangian bias expansion. The central question we address is whether we can find a set of bias parameters that fits both real- and redshift-space $k$NN-CDFs, while also providing a good description of the corresponding $P(k)$ measurements. Since $k$NN-CDFs are sensitive to all $N$-point correlations, including the two-point statistics, agreement with $P(k)$ provides a useful consistency check.
The general methodology for measuring the $k$NN-CDFs on the data and computing their predictions is explained in Section~\ref{sec:knnmodel}, with the implementation details of the hybrid Lagrangian framework provided in Section~\ref{sec:pipeline}.

The real-space $k$NN-CDFs for galaxies are shown in Figure~\ref{fig:Real_LRG_knnCdf_1024p}, where the left panel presents the auto clustering, while the right panel presents the galaxy--matter cross clustering. The 1NN-, 2NN-, and 3NN-CDFs are shown in maroon, green, and golden yellow, respectively. The circular points denote the $k$NN-CDF measured from the galaxy catalogue, while the dashed lines show the corresponding model predictions from the hybrid Lagrangian bias framework. The agreement between the data and model from this figure appears good.
The corresponding residuals between the data and model predictions are shown in Figure~\ref{fig:Real_LRG_knnCdf_1024p_Resid}. The top and bottom rows correspond to the auto and cross clustering, respectively, while the columns from left to right correspond to the 1NN-, 2NN-, and 3NN-CDFs. The solid curves show the residuals for the case when predictions are obtained using the model parameters inferred from the real-space $k$NN-CDFs, while the dashed curves show the same for the case when using the model parameters inferred from the galaxy auto and galaxy--matter cross power spectra in real space. The grey shaded region denotes the scatter in the $k$NN-CDF measurements across 100 realisations, while the shaded region extending up to a given $r$ in each panel marks the scales where the agreement between the data and model does not hold.
We find that the $k$NN-CDFs in the auto clustering case can be well modelled down to $10\,\mathrm{Mpc}/h$, $16\,\mathrm{Mpc}/h$, and $18.5\,\mathrm{Mpc}/h$ for the 1NN-, 2NN-, and 3NN-CDFs, respectively. For the cross clustering between galaxies and the matter field, the corresponding scales are $12\,\mathrm{Mpc}/h$, $14\,\mathrm{Mpc}/h$, and $15\,\mathrm{Mpc}/h$ for the 1NN-, 2NN-, and 3NN-CDFs, respectively.

The $k$NN-CDFs for galaxies in redshift space are shown in Figure~\ref{fig:Rsd_LRG_knnCdf_1024p}, with the auto clustering shown in the left panel and the cross clustering in the right panel. The cross clustering is measured between galaxies in redshift space and the matter field in real space. This setup could be relevant to analyses combining galaxy clustering with weak lensing measurements, where the galaxy field is observed in redshift space while lensing probes the underlying matter distribution (see e.g. \cite{Mandelbaum2013,Blake2016,DESY6,Semenaite2026}).
The panel layout is the same as in the real-space case shown in Figure~\ref{fig:Real_LRG_knnCdf_1024p}, and the agreement between the data and model appears good.
The corresponding residual plots are presented in Figure~\ref{fig:Rsd_LRG_knnCdf_1024p_Resid}, where the solid curves show the residuals for the case when predictions are obtained using the bias parameters inferred from the real-space $k$NN-CDFs, while the dashed curves show the residuals for the case when using bias parameters inferred from the real-space $P(k)$ measurements. The FoG parameters used in the modelling are the ones that are inferred from redshift-space $k$NN-CDFs. As can be seen, the modelling of the $k$NN-CDFs for the galaxy auto clustering is good down to $10\,\mathrm{Mpc}/h$, $16.5\,\mathrm{Mpc}/h$, and $19\,\mathrm{Mpc}/h$ for the 1NN-, 2NN-, and 3NN-CDFs, respectively. For galaxy--matter cross clustering, the corresponding scales are $12\,\mathrm{Mpc}/h$, $16.5\,\mathrm{Mpc}/h$, and $19\,\mathrm{Mpc}/h$ for the 1NN-, 2NN-, and 3NN-CDFs, respectively. The minimum reliable scale for $k$NN-CDFs in both real and redshift space shifts to larger $r$ for higher nearest-neighbour orders, suggesting that the perturbative expansion becomes less accurate in high-density regions, since these orders start probing progressively denser regimes. Including higher-order operators in the bias expansion might improve the modelling in these regimes, albeit at the cost of additional free parameters.

\begin{figure}
    \centering
    \includegraphics[width=0.975\linewidth]{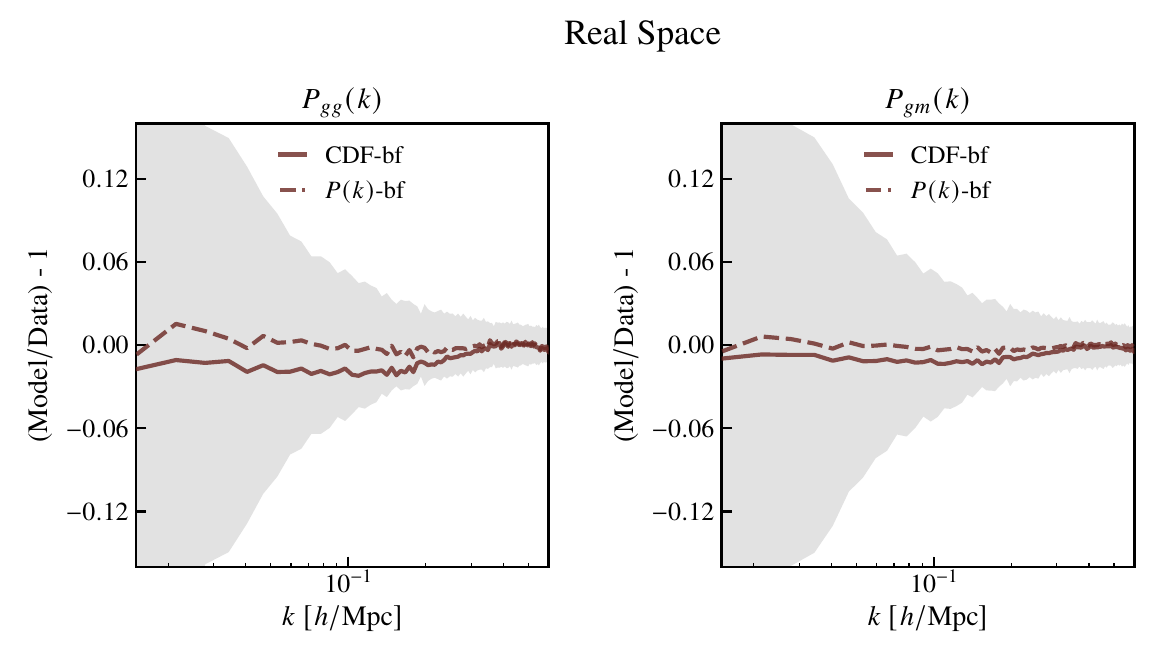}
    \caption{Residuals between the $P(k)$ measurements and model predictions for galaxies in real space. The left panel shows the residuals for the galaxy auto power spectrum $P_{gg}(k)$, while the right panel shows the same for the galaxy--matter cross power spectrum $P_{gm}(k)$. In each panel, the solid line shows the residuals for predictions obtained using the bias parameters inferred from the galaxy $k$NN-CDFs, while the dashed line shows the residuals for predictions obtained using the bias parameters inferred from the real-space galaxy auto and galaxy--matter cross power spectra. The grey shaded regions denote the scatter in the measurements across 100 realisations. The model parameters inferred from $k$NN-CDFs can describe the real-space $P(k)$ for galaxies.}
    \label{fig:Real_LRG_Pk_1024p_Resid}
\end{figure}
\begin{figure}
    \centering
    \includegraphics[width=0.975\linewidth]{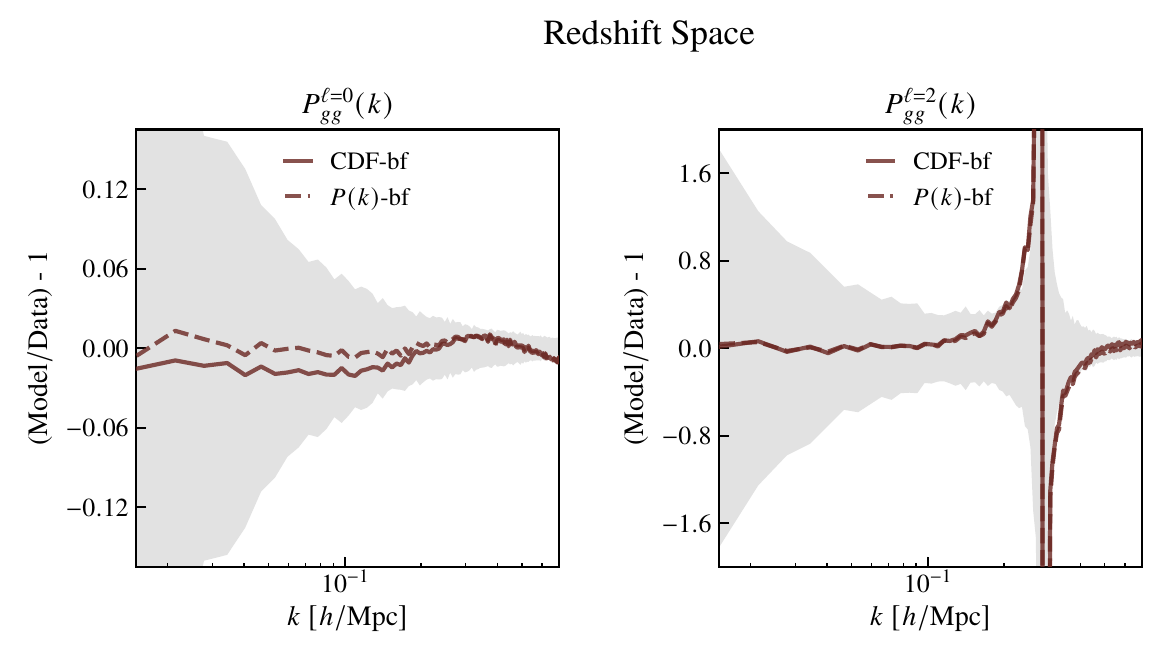}
    \caption{Residuals between the $P(k)$ measurements and model predictions for galaxies in redshift space. The left panel shows the residuals for the monopole $P_{gg}^{\ell=0}(k)$, while the right panel shows the same for the quadrupole $P_{gg}^{\ell=2}(k)$. The line patterns are the same as in Figure~\ref{fig:Real_LRG_Pk_1024p_Resid}. The grey shaded region represents the scatter in the measurements across 100 realisations. The bias parameters inferred from galaxy $k$NN-CDFs provide a good description of the $P(k)$ measurements, whereas the converse does not hold, as shown in Figures~\ref{fig:Real_LRG_knnCdf_1024p_Resid} and \ref{fig:Rsd_LRG_knnCdf_1024p_Resid}.}
    \label{fig:Rsd_LRG_Pk_1024p_Resid}
\end{figure}
\begin{figure}
    \centering
    \includegraphics[width=0.53\linewidth]{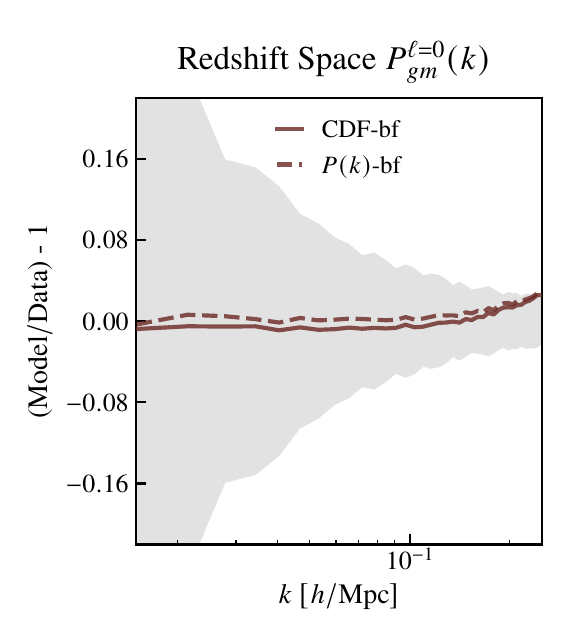}
    \caption{Residuals between the measurement and model prediction for the redshift-space monopole of the galaxy--matter cross power spectrum, $P_{gm}^{\ell=0}$. The galaxies are in redshift space, while the matter field is in real space, as relevant for weak lensing analyses. The line patterns are the same as in Figure~\ref{fig:Real_LRG_Pk_1024p_Resid}, and the grey shaded region represents the scatter in the measurements across 100 realisations. The bias parameters inferred from the $k$NN-CDFs describe the cross power spectrum only up to $k=0.25\,h/\mathrm{Mpc}$, indicating a limitation of the model in this setup.}
\label{fig:Rsd_LRG_Pgm0_1024p_Resid}
\end{figure}

We also find that the bias parameters inferred from galaxy $k$NN-CDFs provide a good description of the corresponding $P(k)$ measurements. Figure~\ref{fig:Real_LRG_Pk_1024p_Resid} shows the real-space residuals, with $P_{gg}(k)$ in the left panel and $P_{gm}(k)$ in the right panel. The solid and dashed curves show the residuals obtained using the model parameters inferred from the $k$NN-CDFs and $P(k)$ measurements, respectively. The grey shaded region denotes the scatter in the corresponding $P(k)$ measurements across 100 realisations.
Figure~\ref{fig:Rsd_LRG_Pk_1024p_Resid} presents the $P(k)$ residuals for galaxies in redshift space. The left panel shows the residuals for the monopole of the galaxy auto power spectrum $P_{gg}^{\ell=0}(k)$, while the right panel shows those for the corresponding quadrupole $P_{gg}^{\ell=2}(k)$. The line convention is the same as in Figure~\ref{fig:Real_LRG_Pk_1024p_Resid}. These results show that for galaxy auto clustering in both real and redshift space, the $P(k)$ residuals obtained using parameters inferred from the $k$NN-CDFs remain within the measurement scatter up to $k=0.6\,h/\mathrm{Mpc}$. Figure~\ref{fig:Rsd_LRG_Pgm0_1024p_Resid} presents the corresponding residuals for the monopole of the galaxy--matter cross power spectrum, $P_{gm}^{\ell=0}(k)$, with galaxies in redshift space and matter in real space. The line convention follows from Figure~\ref{fig:Real_LRG_Pk_1024p_Resid}, and we show the residuals only up to $k=0.25\,h/\mathrm{Mpc}$, beyond which the model no longer describes the data. This indicates a limitation of the redshift-space model in this setup, the investigation of which is beyond the scope of this study and is deferred to future work.

These results demonstrate that the hybrid Lagrangian bias expansion framework can model the $k$NN-CDFs of galaxies in both real and redshift space, and that the model parameters inferred from the $k$NN-CDFs also provide a good description of the $P(k)$ measurements, whereas the converse need not hold. This aligns with the fact that higher-order statistics are more informative and hence more restrictive, unlike the two-point statistics that can allow for a wider parameter range.

\section{Fisher Analysis}\label{sec:Fisher}
\begin{figure*}
    \centering
    \includegraphics[width=0.935\linewidth]{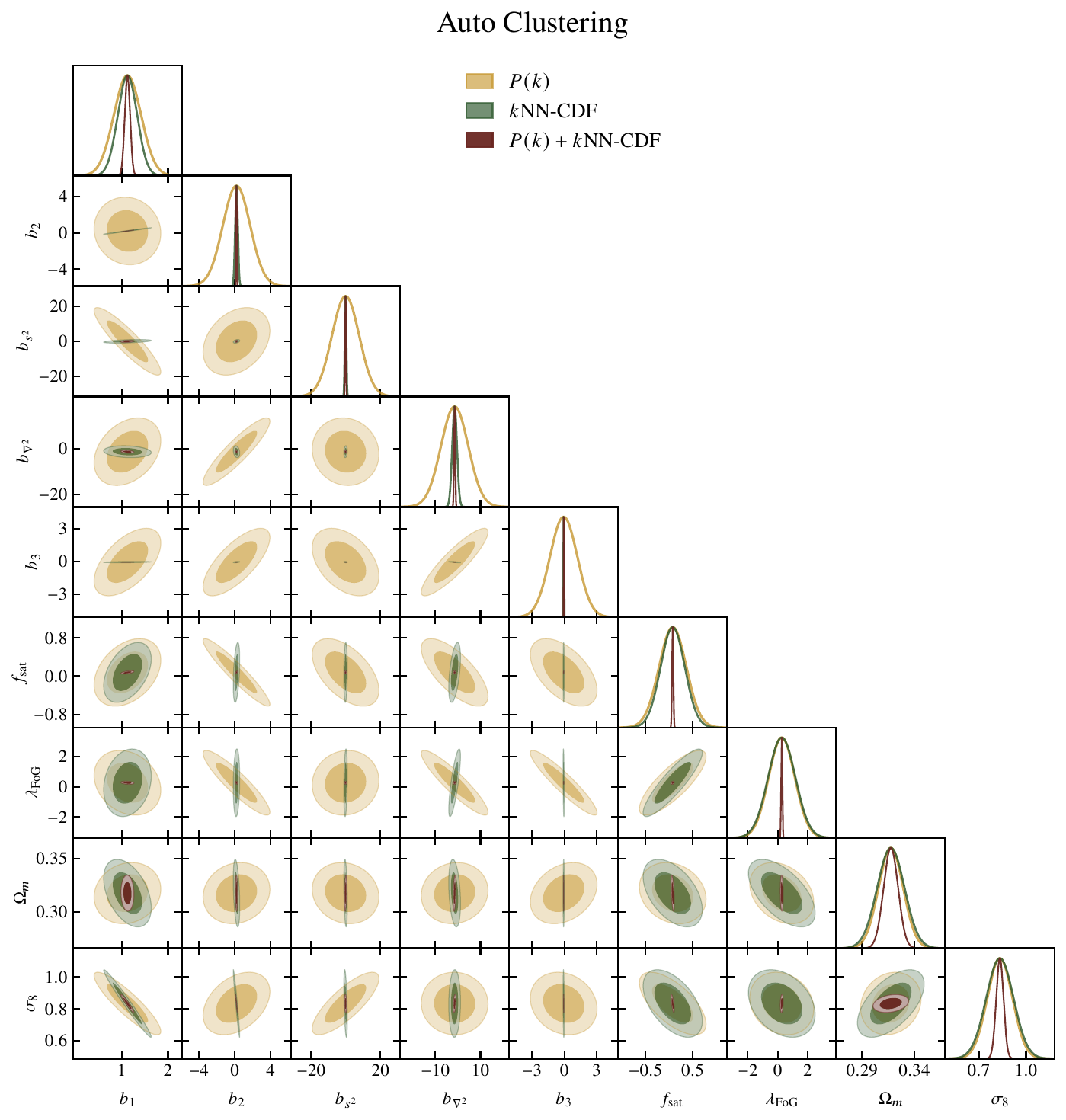}
    \caption{Fisher constraints on bias ($b_1$, $b_2$, $b_{s^2}$, $b_{\nabla^2}$, $b_3$), FoG ($f_{\rm sat}$, $\lambda_{\rm FoG}$), and cosmological ($\Omega_m$, $\sigma_8$) parameters, based on redshift-space auto clustering of galaxies (Figure~\ref{fig:GalxHOD_512p} shows the corresponding HOD). The off-diagonal panels show the $1\sigma$ and $2\sigma$ marginalised 2D posterior contours, while diagonal panels show the marginalised 1D posterior distributions. The constraints shown in golden yellow correspond to $P(k)$ (monopole+quadrupole, across $0.015\,h/\mathrm{Mpc} \leq \! k \!\leq 0.5\,h/\mathrm{Mpc}$; those in green to the $k$NN-CDFs, including the first three nearest neighbours (over $r=16$, $14$, and $20$ up to $30\,\mathrm{Mpc}/h$ for the 1NN-, 2NN-, and 3NN-CDFs, respectively); and those in maroon to the combination of $P(k)$ and the $k$NN-CDFs. The corresponding marginalised $1\sigma$ uncertainties are listed in Table~\ref{tab:FisherConstraints}. The $k$NN-CDFs provide tighter constraints on all the bias parameters, particularly the higher-order ones, while the constraints on the other parameters are comparable to those from $P(k)$. The two statistics are complementary across the scales considered, which makes their combined analysis the most effective for constraining the parameters. For several parameter pairs, the $P(k)$ and $k$NN-CDF contours have different degeneracy directions; combining them therefore tightens the constraints on all the parameters considered relative to either statistic alone. Relative to $P(k)$ alone, the combination improves $b_3$ the most, while the constraints on $\sigma_8$ and $\Omega_m$ improve by factors of 3.5 and 1.7, respectively.}
\label{fig:Fisher_AutoClstr_Main}
\end{figure*}
\begin{figure*}
    \centering
    \includegraphics[width=0.935\linewidth]{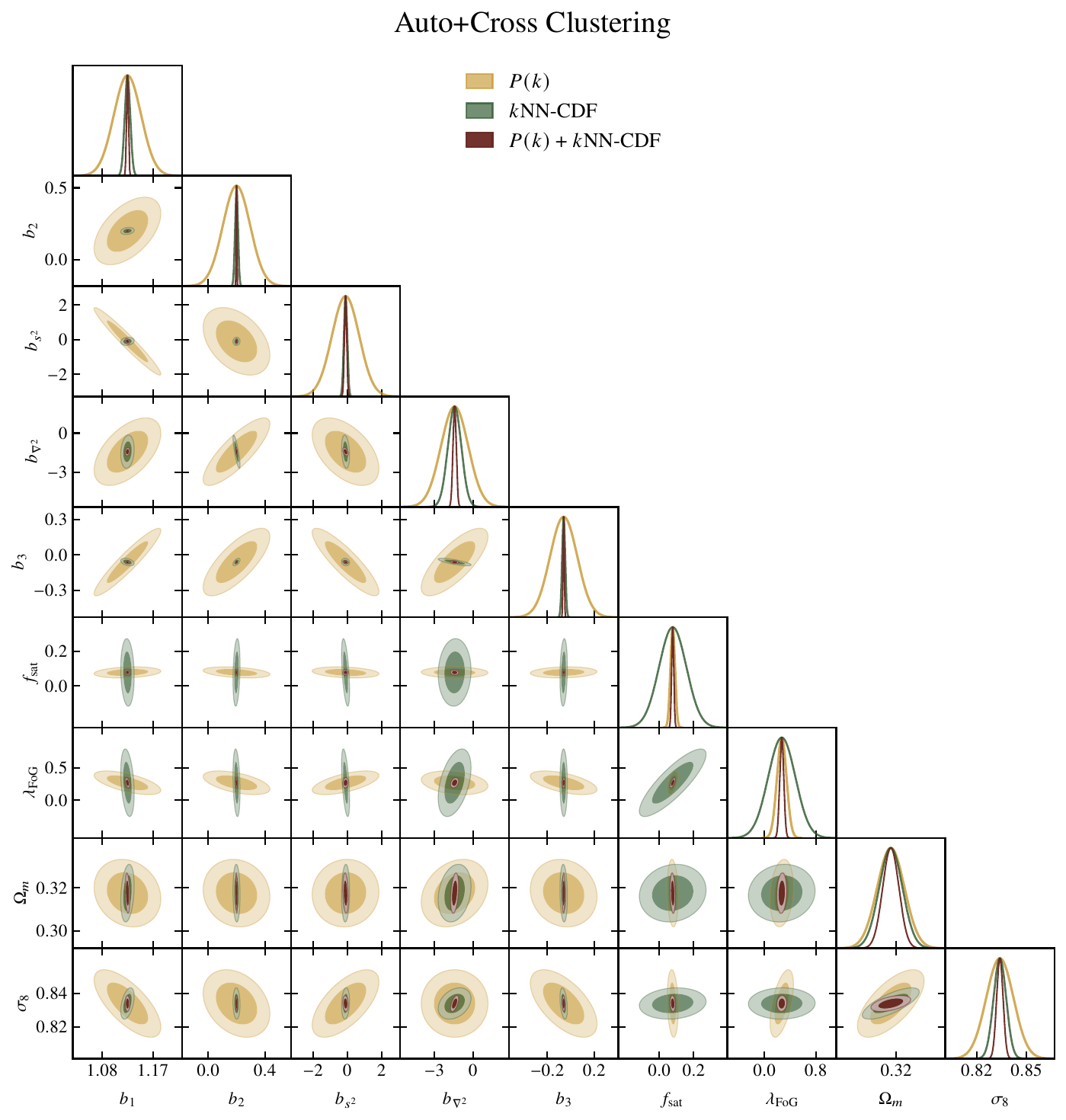}
     \caption{Fisher constraints on bias ($b_1$, $b_2$, $b_{s^2}$, $b_{\nabla^2}$, $b_3$), FoG ($f_{\rm sat}$, $\lambda_{\rm FoG}$), and cosmological ($\Omega_m$, $\sigma_8$) parameters, from the combination of galaxy auto and galaxy--matter cross clustering (galaxies in redshift space and matter field in real space; refer to the text for details). The colour convention and panel layout are the same as in Figure~\ref{fig:Fisher_AutoClstr_Main}. The scales used for $P_{gg}(k)$ and the $k$NN-CDFs are the same as in the auto clustering case, while $P_{gm}^{\ell=0}(k)$ is included up to $k=0.25\,h/\mathrm{Mpc}$, owing to the limited validity of the model for this statistic at smaller scales. The corresponding marginalised $1\sigma$ uncertainties are listed in Table~\ref{tab:FisherConstraints}. The $k$NN-CDFs provide tighter constraints than $P(k)$ on all the bias and cosmological parameters, while $P(k)$ is better at constraining the FoG parameters. Combining $P(k)$ and the $k$NN-CDFs in this case yields the tightest constraints across all Fisher forecasts considered in our work. In particular, adding the $k$NN-CDFs to $P(k)$ improves the constraints on $b_2$ the most compared to $P(k)$ alone, while the constraints on $\sigma_8$ and $\Omega_m$ improve by factors of 4.2 and 1.7, respectively.}
    \label{fig:Fisher_AutoCross_Main}
\end{figure*}
\begin{table*}
\centering
\caption{Marginalised $1\sigma$ constraints on the bias ($b_1, b_2, b_{s^2}, b_{\nabla^2}, b_3$), FoG ($f_{\rm sat}$, $\lambda_{\rm FoG}$), and cosmological ($\Omega_m$, $\sigma_8$) parameters obtained from the Fisher forecasts based on redshift-space galaxies. The table presents constraints from $P(k)$, the $k$NN-CDFs, and their combination for (i) galaxy auto clustering (Figure~\ref{fig:Fisher_AutoClstr_Main}) and (ii) combination of galaxy auto clustering and galaxy--matter cross clustering (matter field in real space; Figure~\ref{fig:Fisher_AutoCross_Main}).}
\label{tab:FisherConstraints}
\begin{tabular}{c @{\hspace{0.5cm}} ccc @{\hspace{0.75cm}} ccc}
\toprule
\textbf{Parameters}
& \multicolumn{3}{c}{\textbf{Auto Clustering}}
& \multicolumn{3}{c}{\textbf{Auto + Cross Clustering}} \\
& $\sigma_{P(k)}$ & $\sigma_{k\text{NN-CDF}}$ & $\sigma_{P(k)+k\text{NN-CDF}}$
& $\sigma_{P(k)}$ & $\sigma_{k\text{NN-CDF}}$ & $\sigma_{P(k)+k\text{NN-CDF}}$ \\
\midrule
$b_1$               & 0.294  & 0.210  & 0.0569  & 0.0242  & 0.00487 & 0.00169 \\
$b_2$               & 1.52   & 0.146  & 0.0391  & 0.0958  & 0.0101  & 0.00246 \\
$b_{s^2}$           & 7.86   & 0.409  & 0.266   & 0.796   & 0.0956  & 0.0666  \\
$b_{\nabla^2}$      & 6.03   & 1.06   & 0.254   & 1.06    & 0.520   & 0.127   \\
$b_3$               & 1.26   & 0.0225 & 0.00797 & 0.116   & 0.0129  & 0.00511 \\
$f_{\rm sat}$       & 0.288  & 0.256  & 0.0133  & 0.0130  & 0.0792  & 0.00619 \\
$\lambda_{\rm FoG}$ & 0.874  & 0.912  & 0.0390  & 0.0744  & 0.212   & 0.0313  \\
$\Omega_m$          & 0.0118 & 0.0130 & 0.00694 & 0.00640 & 0.00546 & 0.00381 \\
$\sigma_8$          & 0.0802 & 0.0866 & 0.0230  & 0.00822 & 0.00381 & 0.00194 \\
\bottomrule
\end{tabular}
\end{table*}

In this section, we present the Fisher analysis performed to quantify the constraining power of the $k$NN-CDFs, the $P(k)$, and their combination on the bias parameters of the hybrid Lagrangian bias expansion, the FoG parameters of the RSD model, and the two cosmological parameters $\Omega_m$ and $\sigma_8$.
As discussed in Section~\ref{sec:pipeline}, the two statistics access different aspects of the clustering information. The $P(k)$ quadrupole is expected to be the most sensitive to the FoG parameters, as it captures the anisotropic clustering information imprinted by RSDs, which is not captured by the $k$NN-CDFs or the $P(k)$ monopole. On the other hand, $k$NN-CDFs, being sensitive to higher-order correlations, are expected to be more sensitive to the higher-order bias parameters, to which $P(k)$ has limited sensitivity, while the larger scales probed by $P(k)$ are expected to carry additional information on some of the cosmological parameters. We therefore expect combining the two statistics to be the most effective approach.

Our Fisher analysis is based on galaxies constructed from the \textit{Quijote} simulation suite, with a box size of $L=1\,\mathrm{Gpc}/h$, consisting of $512^3$ CDM particles, and adopts a fiducial cosmology (see Section~\ref{sec:sim_tracers} for cosmology details).
The galaxies are obtained using the HOD prescription given by equations~\ref{HOD_Cen} and~\ref{HOD_Sat}. We adopt a minimum halo mass of $1.31\times10^{13}\, M_{\odot}/h$, corresponding to 20 CDM particles, as a conservative resolution threshold. The particle resolution in this particular simulation suite restricts the resolved halo population to relatively high masses; we therefore tailor the HOD parameters to this halo mass range rather than adopting values calibrated for lower mass halos. Including halos from the upper end of the mass range leads to a discrepancy between the $P(k)$ data and the model predictions obtained using the bias parameters inferred from the $k$NN-CDFs, suggesting that the current bias expansion becomes less adequate for this highly biased tracer population. We therefore impose an upper halo mass cut of $10^{14}\, M_{\odot}/h$ before populating with the HOD. In the case of high-resolution simulations, where the halo sample consists of a larger population of lower-mass halos, we obtain satisfactory agreement without requiring such a mass cut (as shown in Section~\ref{sec:ModellingResults}).

The following values are used for the HOD parameters: $\log M_{\rm min}=13.2$, $\sigma_{\log M}=0.075$, $\log M_0=13.64$, $\log M_1=13.64$, and $\alpha=1.35$. While this HOD is not intended to reproduce a very realistic galaxy sample, it provides a controlled setup for carrying out the Fisher forecasts. Figure~\ref{fig:GalxHOD_512p} shows the HOD form for this galaxy sample and Appendix~\ref{sec:Fisher_Appendix} presents the relevant residual plots for the $k$NN-CDFs and $P(k)$, comparing the data with the model predictions.

The Fisher forecast formalism provides an estimate of the parameter constraints achievable from a given set of observables (see e.g. \cite{Tegmark1997}). Assuming that the likelihood for the observables can be approximated by a Gaussian, the Fisher matrix corresponds to the curvature of the log-likelihood around the fiducial parameter values, with its entries given by
\eq{FisherMatrix}{
F_{ij} = \sum_{m,n} \frac{\partial \mathbf{D}_m}{\partial \theta_i} \,\left[\mathbf{C}^{-1}\right]_{mn}\, \frac{\partial \mathbf{D}_n}{\partial \theta_j} \,,
}
where $\mathbf{D}$ denotes the data vector, which, depending on the analysis, consists of the $k$NN-CDF predictions, the $P(k)$ predictions, or a concatenation of the two for the joint Fisher analysis, and $\mathbf{C}$ denotes the covariance matrix. The marginalised $1\sigma$ constraint on each parameter is given by
\eq{FisherError}{
\sigma(\theta_i) = \sqrt{(F^{-1})_{ii}}\, \,.
}

To estimate the derivatives, we use the central difference formula:
\eq{Derivative}{
\frac{\partial \mathbf{D}}{\partial \theta_i} =
\frac{\mathbf{D}(\theta_i+\Delta\theta_i) - \mathbf{D}(\theta_i-\Delta\theta_i)}
{2\Delta\theta_i}\,.
}

The parameter space of the model for this galaxy sample includes five bias parameters, $b_1$, $b_2$, $b_{s^2}$, $b_{\nabla^2}$, and $b_3$, along with two FoG parameters, $f_{\rm sat}$ and $\lambda_{\rm FoG}$. The $P_{gg}^{\ell=0}(k)$ model includes one additional parameter controlling the scale-independent stochastic contribution. For the Fisher forecast, we use the model parameter values inferred from $k$NN-CDFs as the fiducial set. These parameters also provide a good description of redshift-space $P(k)$ measurements, thus enabling a Fisher analysis combining the two statistics. It is worth noting that the inferred $f_{\rm sat}$ for this galaxy sample is very close to the true satellite fraction, indicating that the FoG model here captures the satellite contribution adequately. However, for the LRG-like galaxy sample, whose modelling results were presented in Section~\ref{sec:ModellingResults}, the inferred $f_{\rm sat}$ differs from the true value by about 0.03, suggesting that the FoG parameters serve more as an effective description in this case.

The covariance matrices for the $P(k)$ measurements, the $k$NN-CDFs, and their joint combination are computed using the model predictions evaluated at the fiducial parameter values. We use 1000 realisations of the fiducial cosmology for these computations. The Hartlap correction factor is applied to $\mathbf{C}^{-1}$ to account for the finite number of realisations used to estimate the covariance matrices \citep{Hartlap2007}. While checking the convergence of the derivatives, we find that those with respect to the cosmological parameters converge with 100 realisations (to which we had access) for each of the $\Omega_m^{\pm}$ and $\sigma_8^{\pm}$ cosmologies, whereas those with respect to some of the model parameters require 1000 realisations to converge. We therefore use the respective numbers of realisations for the derivatives. The step size $\Delta\theta_i$ for computing the derivatives with respect to model parameters is set to $5\%$ of the corresponding parameter value.

The Fisher constraints on the bias, FoG, and cosmological parameters from the auto clustering of galaxies in redshift space are shown in Figure~\ref{fig:Fisher_AutoClstr_Main}. The $P(k)$ analysis includes the monopole and quadrupole over the range $0.015\,h/\mathrm{Mpc} \leq \! k \!\leq 0.5\,h/\mathrm{Mpc}$, with the corresponding constraints shown in golden yellow. For the $k$NN-CDFs, we include the first three nearest-neighbours over scales from $r=16\,\mathrm{Mpc}/h$, $14\,\mathrm{Mpc}/h$, and $20\,\mathrm{Mpc}/h$ for the 1NN-, 2NN-, and 3NN-CDFs, respectively, to $r=30\,\mathrm{Mpc}/h$ in each case. The corresponding constraints are shown in green. The scales chosen for each statistic are those over which the model can describe the data (refer to Figures~\ref{fig:RsdAutoCross_knnCdf_512p_Resid} and ~\ref{fig:RsdAuto_Pk_512p_Resid}).

Looking at the individual parameter constraints from the two statistics, the $k$NN-CDFs provide tighter constraints on all the bias parameters, particularly the higher-order ones. The constraints on the other parameters are comparable between the $P(k)$ and the $k$NN-CDFs. For $P(k)$ analysis, including the quadrupole alongside the monopole tightens constraints on all parameters, most significantly on the FoG parameters (Figure~\ref{fig:Fisher_AutoClstr_Apndx}).
As discussed before, the two statistics, $P(k)$ and the $k$NN-CDFs, are sensitive to complementary clustering information of the tracers across the scales considered, making their combination the most effective strategy. This combination, shown in maroon in the figure, leads to tighter constraints on all bias, FoG, and cosmological parameters than either statistic alone. Compared to the constraints from $P(k)$, the largest gain from this combination is achieved for $b_3$, while the constraints on the cosmological parameters $\sigma_8$ and $\Omega_m$ improve by factors of 3.5 and 1.7, respectively.

We also perform a Fisher analysis combining the galaxy auto clustering with the galaxy--matter cross clustering. This setup involves the galaxies in redshift space, whereas the matter field is in real space, which could be relevant for weak lensing analyses (as noted in Section~\ref{sec:ModellingResults}).
The scales used for the $k$NN-CDFs and the $P_{gg}(k)$ multipoles ($\ell=0,2$) remain the same as in the auto clustering analysis. However, because the agreement between the data and the model for $P_{gm}^{\ell=0}(k)$ is good only up to $k=0.25\,h/\mathrm{Mpc}$ (Figure~\ref{fig:RsdPgm0_512p_Resid}), we restrict our analysis to this scale for this statistic. Figure~\ref{fig:Fisher_AutoCross_Main} presents the corresponding Fisher forecasts, where the constraints shown in golden yellow correspond to $P(k)$, whereas those shown in green correspond to the $k$NN-CDFs, and finally their combined constraints are shown in maroon. Including the cross clustering alongside the auto clustering improves the constraints for both statistics. Furthermore, combining $P(k)$ and the $k$NN-CDFs in this configuration leads to the tightest overall constraints. Relative to $P(k)$ constraints, the largest gain from this combination is achieved for $b_2$, while the constraints on $\sigma_8$ and $\Omega_m$ improve by factors of 4.2 and 1.7, respectively. The gains for individual statistics from adding the cross clustering information are shown in Figures~\ref{fig:Fisher_Pk_AutoCrossClstr_Apndx} and \ref{fig:Fisher_knnCdf_AutoCrossClstr_Apndx} for $P(k)$ and $k$NN-CDFs, respectively.

\section{Conclusions and Future Outlook}\label{sec:Conclusions}

In this work, we have developed a pipeline to model the $k$NN-CDFs of biased tracers in redshift space. The modelling pipeline is based on the hybrid Lagrangian bias expansion approach for the tracer--matter connection, along with a model for RSDs including the central distortion and FoG effects. The sensitivity of the $k$NN-CDFs to all $N$-point correlations, together with the computational ease of measuring them on data, makes them a compelling statistic in the current era of precision cosmology. Thus, modelling the $k$NN-CDFs for biased tracers of interest is crucial to harness their statistical gains.

Our main modelling results are based on a galaxy sample constructed using an HOD to mimic a typical LRG-like sample at $z=0.5$, using the high-resolution \textit{Quijote} simulations which include $1024^3$ CDM particles. We demonstrate that the model can describe the measured $k$NN-CDFs for auto clustering of galaxies in redshift space as well as galaxy--matter cross clustering over a quasi-linear to linear range of scales. The agreement between the data and the model at best reaches down to $r=10\,\mathrm{Mpc}/h$ for the 1NN-CDFs in auto clustering. 
The minimum scale where the data and model agree shifts to relatively larger scales for higher nearest-neighbour orders, which implies that the model has limited validity in high-density regions, as higher-order $k$NN-CDFs probe progressively denser regimes. Including higher-order operators in the bias expansion might improve the modelling in this regime, albeit at the cost of introducing additional bias parameters. A systematic study of including higher nearest-neighbour orders in the $k$NN-CDFs, by extending the bias operators basis, is deferred to future work.

Another important result of our analysis is that the model parameters inferred from the $k$NN-CDFs also provide a good description of the galaxy $P(k)$ in both real and redshift space, indicating the feasibility of combining the two statistics for joint parameter inference. This result also supports the model's validity, as it is able to describe both statistics in real as well as redshift space. However, the best-fit parameters inferred from $P(k)$ do not provide an accurate description of the $k$NN-CDFs. This aligns with the fact that the $k$NN-CDFs are more informative and thus impose tighter restrictions on the model parameters, whereas the two-point statistics can allow a broader range of parameter values.

We also performed a Fisher forecast analysis to quantify the constraints on the bias parameters ($b_1, \,b_2,\, b_{s^2},\, b_{\nabla^2},\, b_3$), the FoG parameters ($f_{\rm sat}$,  $\lambda_{\rm FoG}$), and the cosmological parameters ($\Omega_m$, $\sigma_8$), coming from the $k$NN-CDFs, the $P(k)$, and their combination. The analysis is based on galaxies constructed using HOD at $z=0.5$ from the \textit{Quijote} simulations, which contain $512^3$ CDM particles. The model parameters inferred from the $k$NN-CDFs also describe $P(k)$, and therefore this parameter set is adopted as the fiducial set around which derivatives with respect to the model parameters are computed.
For the auto clustering of galaxies in redshift space, we find that the $k$NN-CDFs provide tighter constraints on the bias parameters than $P(k)$, while the constraints on the FoG and cosmological parameters are comparable between the two statistics. Although the $k$NN-CDFs are not sensitive to the angular dependence of clustering probed by the quadrupole, they still provide appreciable constraints on the FoG parameters. Combining $P(k)$ and the $k$NN-CDFs yields the tightest constraints for auto clustering case on all the parameters considered. Compared to $P(k)$ alone, the combination improves the constraints on $\sigma_8$ and $\Omega_m$ by factors of 3.5 and 1.7, respectively, whereas the largest gain is obtained for $b_3$.
Including the galaxy--matter cross clustering alongside the auto clustering for the Fisher analysis improves the constraints for both statistics. In these forecasts, the $k$NN-CDFs provide tighter constraints on the bias and cosmological parameters, whereas $P(k)$ has better constraining power on the FoG parameters. Combining $P(k)$ and the $k$NN-CDFs yields the tightest constraints across all the cases considered. Compared to $P(k)$ alone, the largest improvement for this combined case is obtained for $b_2$, while the constraints on $\sigma_8$ and $\Omega_m$ improve by factors of 4.2 and 1.7, respectively.

These results demonstrate that the redshift-space $k$NN-CDFs of biased tracers of matter can be successfully modelled within the hybrid Lagrangian bias expansion framework. They also highlight the potential of combining the $k$NN-CDFs with conventional two-point statistics to obtain tighter constraints on cosmological and tracer--matter model parameters.
A natural step to take next is to apply this modelling pipeline to the $k$NN-CDFs of observed galaxies. However, cosmological inference using these statistics, for instance through Markov Chain Monte Carlo (MCMC) analyses, would require a faster prediction pipeline. To this end, we are working towards developing an emulator for the $k$NN-CDFs as a function of the bias, FoG, and cosmological parameters. This would facilitate the application of the framework to observational data.

In future work, we plan to extend the formalism to incorporate the angular dependence of the $k$NN-CDFs, which would provide additional information for clustering analyses in redshift space. 
We also intend to explore simulation-based inference (SBI; \cite{Cranmer2020,Alsing2019}) in the context of $k$NN-CDFs. SBI enables parameter inference using forward simulations without requiring an explicit evaluation of the likelihood. This is particularly relevant when the likelihood of the data vector is intractable or when commonly used likelihood approximations, such as a multivariate Gaussian, may not adequately capture its distribution. Depending on the method, SBI can learn the posterior distribution directly or estimate quantities needed to perform inference from simulated data and their corresponding parameter values. Once trained and validated, an inference model can enable efficient posterior estimation, making SBI a promising approach for extending the use of higher-order statistics such as $k$NN-CDFs to cosmological inference. 
The hybrid Lagrangian bias framework, though very promising, has certain limitations as well, one of which is that the advected tracer overdensity field can take values below $-1$ in some regions, which is unphysical. The Lagrangian bias field framework recently introduced in \cite{Banerjee2026} bypasses this limitation by describing Lagrangian bias as a property of a Gaussian random field rather than solely through a set of bias coefficients inferred from data. It would thus be interesting to explore modelling the $k$NN-CDFs using this new approach.

\section*{Acknowledgements}
EC acknowledges the Ministry of Education, Government of India, and the Indian Institute of Science Education and Research, Pune, for the support of the Prime Minister Research Fellowship (PMRF). AB’s work was partially supported by the Startup Research Grant (SRG/2023/000378) from the Anusandhan National Research Foundation (ANRF), India (previously SERB, India). We also acknowledge the PARAM Brahma Facility under the National Supercomputing Mission, Government of India, at the Indian Institute of Science Education and Research, Pune, for providing the computing resources for this work.

\section*{Data Availability}
The Quijote simulations used in this work are publicly available and can be accessed from \url{https://quijote-simulations.readthedocs.io/}. The codes and other datasets used in this work can be made available upon reasonable request.

\bibliographystyle{mnras}
\bibliography{ref}
\appendix
\section{Modelling Results for Halos}\label{sec:HaloResults}
\begin{figure*}
    \centering
    \includegraphics[width=0.85\linewidth]{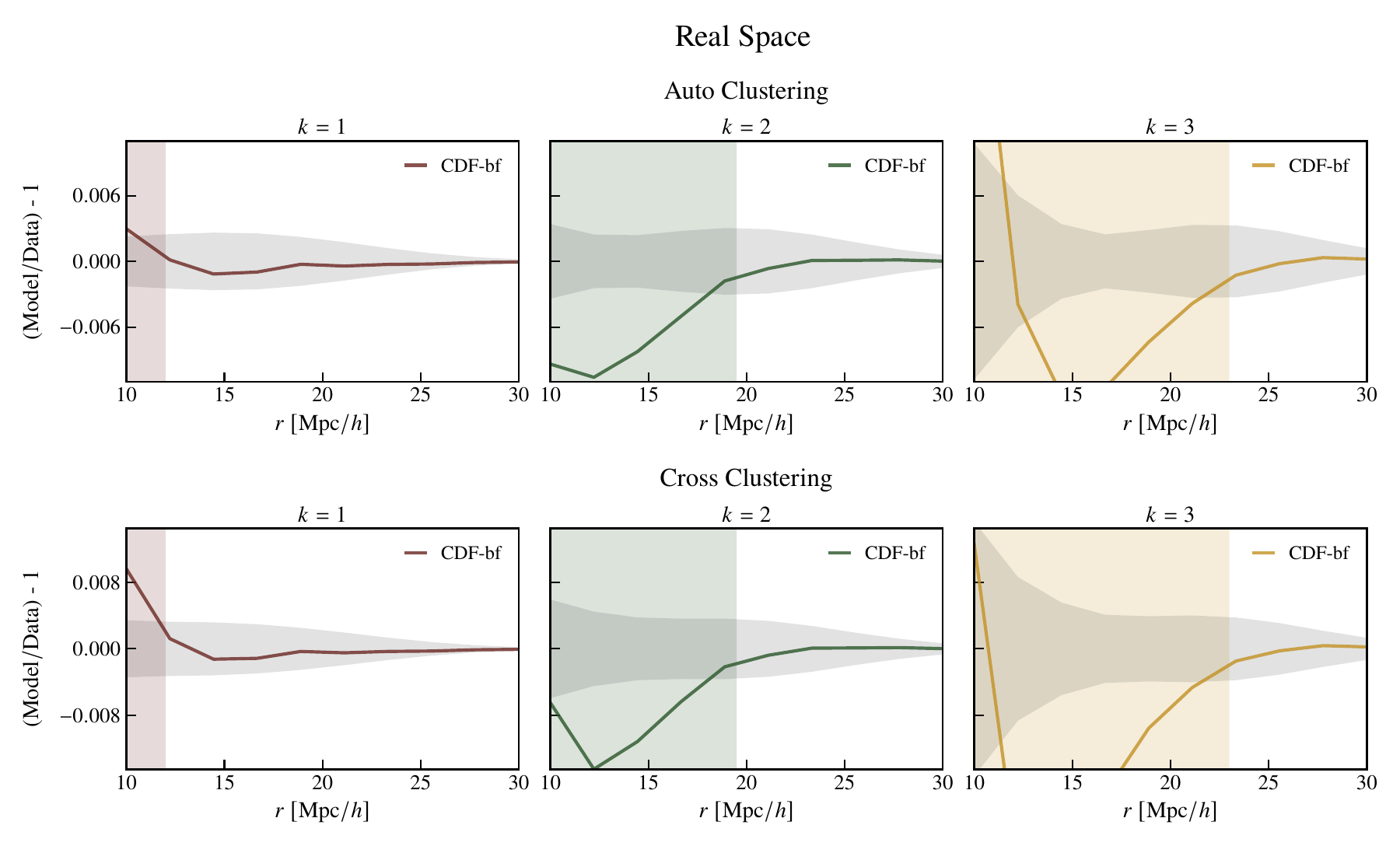}
    \caption{Residuals between the $k$NN-CDF measurements of halos and model predictions within the hybrid Lagrangian bias modelling framework. The top row presents the halo auto clustering case, while the bottom row presents the halo--matter cross clustering case. The columns from left to right show the 1NN-, 2NN-, and 3NN-CDFs, and the grey shaded region denotes the scatter in measurements across 100 realisations. The shaded region extending up to a given $r$ in each panel marks the scales where the agreement between data and model does not hold. The agreement between data and model here at best reaches $r=12\,\mathrm{Mpc}/h$ for 1NN-CDFs in both auto and cross clustering.}
    \label{fig:RealHalos_knnCdf_1024p_Resid}
\end{figure*}
\begin{figure*}
    \centering
    \includegraphics[width=0.85\linewidth]{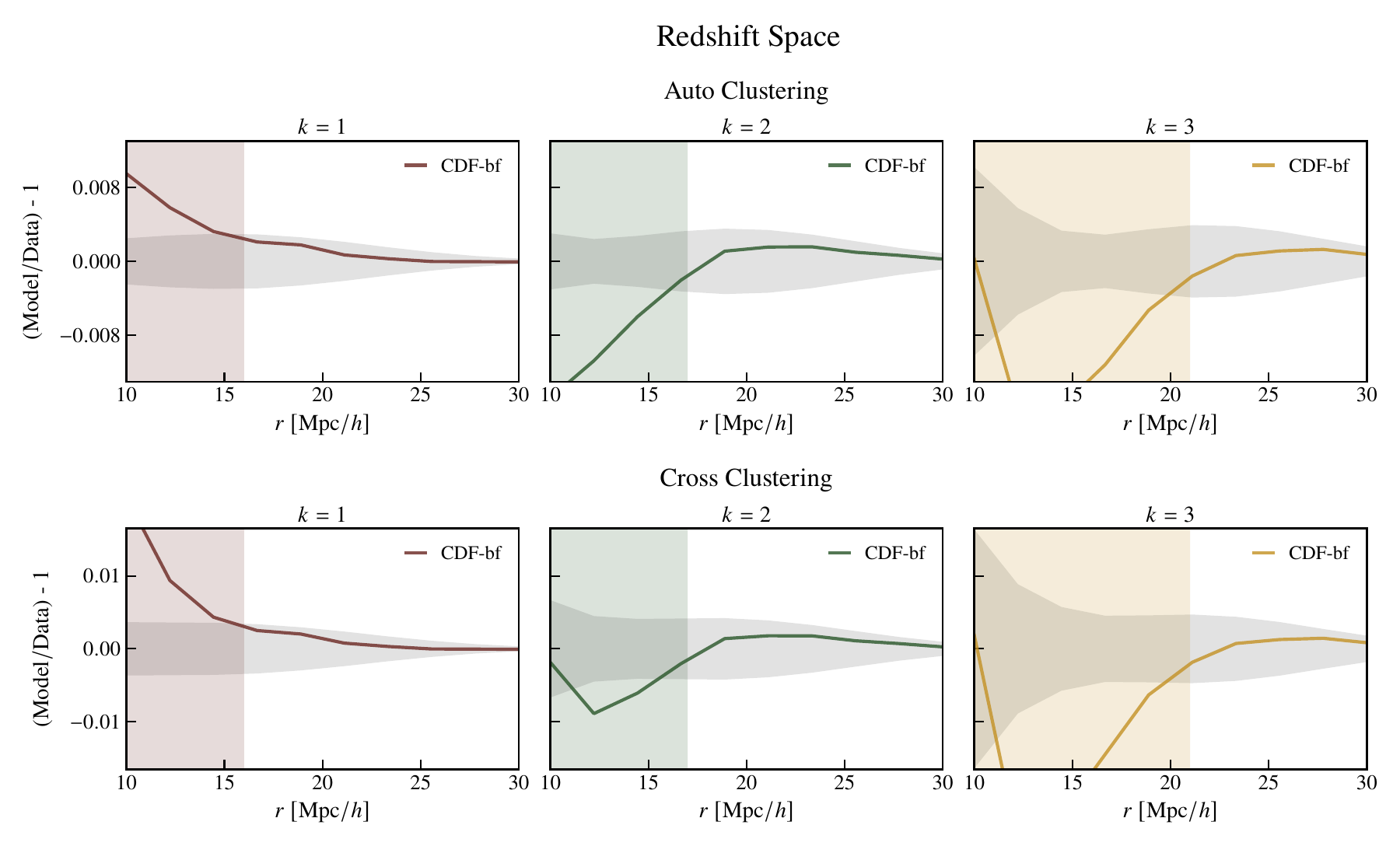}
    \caption{Residuals between the measured and predicted $k$NN-CDFs for halos in redshift space. The top row shows halo auto clustering, while the bottom row shows halo--matter cross clustering, with the matter field in real space. The columns from left to right show the 1NN-, 2NN-, and 3NN-CDFs, and the grey shaded region denotes the scatter in measurements across 100 realisations. The model has limited validity on small scales in this case, with the best agreement for the 1NN-CDF extending to $r=16\,\mathrm{Mpc}/h$.}
    \label{fig:RsdHalos_knnCdf_1024p_Resid}
\end{figure*}
\begin{figure}
    \centering
    \includegraphics[width=0.98\linewidth]{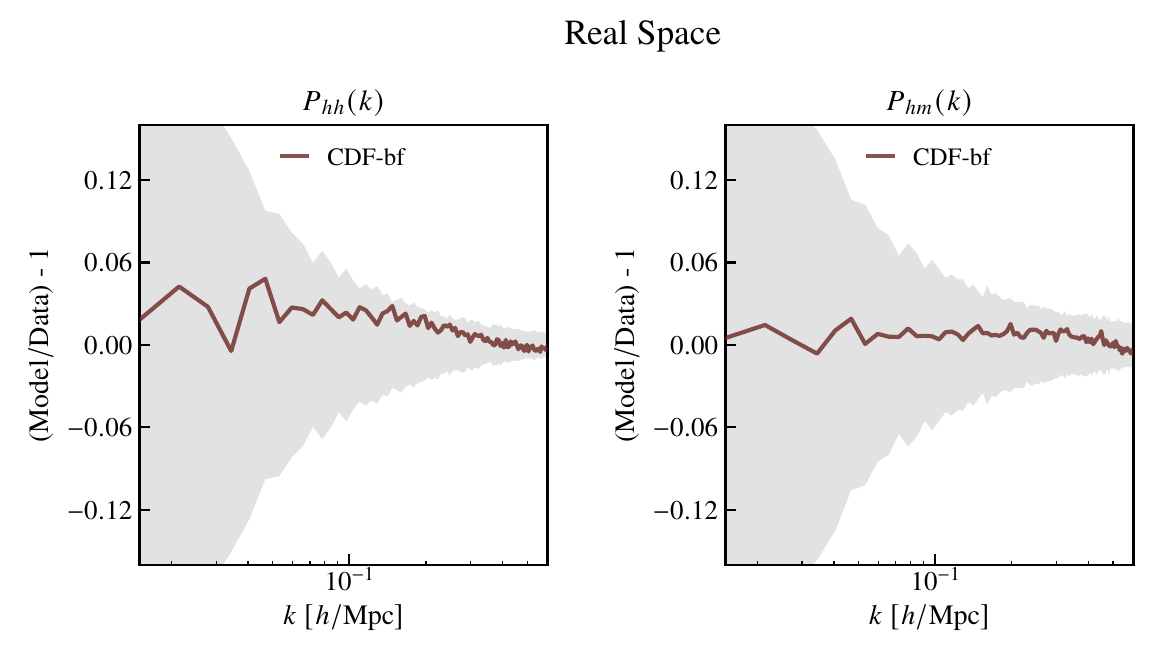}
    \caption{Residuals between the measured and predicted $P(k)$ for halos in real space, using the bias parameters inferred from the $k$NN-CDFs. The left panel shows the halo auto power spectrum $P_{hh}(k)$, while the right panel shows the halo--matter cross power spectrum $P_{hm}(k)$. The grey shaded region represents the scatter in measurements across 100 realisations. The bias parameters inferred from the halo $k$NN-CDFs can describe the $P(k)$ measurements up to $k=0.6\,h/\mathrm{Mpc}$.}
    \label{fig:RealHalos_Pk_1024p_Resid}
\end{figure}
\begin{figure}
    \centering
    \includegraphics[width=0.98\linewidth]{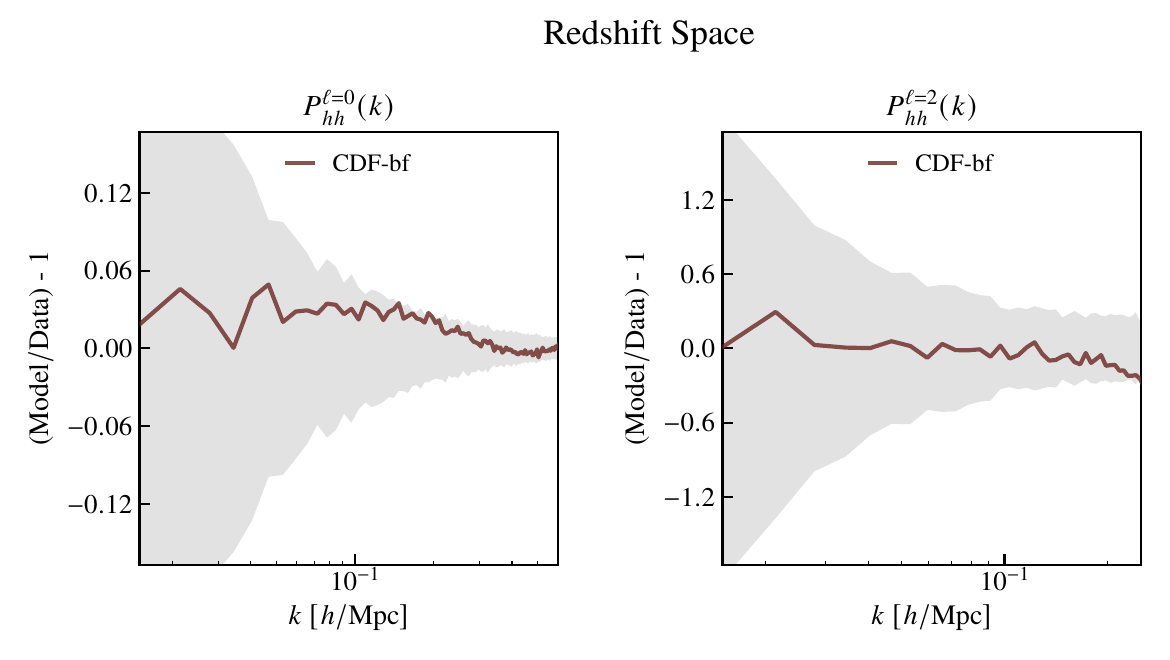}
    \caption{Residuals between the measured and predicted $P(k)$ for halos in redshift space, using the bias parameters inferred from the $k$NN-CDFs. The left panel shows the monopole of the halo auto power spectrum $P_{hh}^{\ell=0}(k)$, while the right panel shows the corresponding quadrupole $P_{hh}^{\ell=2}(k)$. The grey shaded region represents the scatter in measurements across 100 realisations. The bias parameters inferred from the halo $k$NN-CDFs are able to describe the monopole up to $k=0.6\,h/\mathrm{Mpc}$, while the agreement for the quadrupole is limited to $k=0.25\,h/\mathrm{Mpc}$.}
    \label{fig:RsdHalos_Pk_1024p_Resid}
\end{figure}
\begin{figure}
    \centering
    \includegraphics[width=0.55\linewidth]{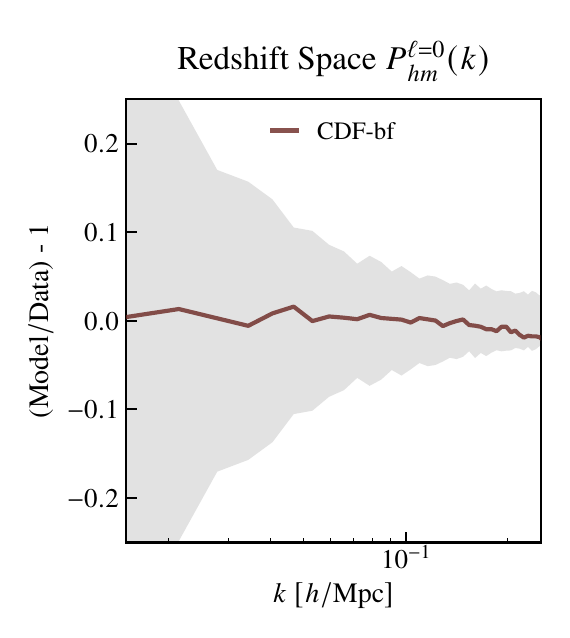}
    \caption{Residual between the measurement and model prediction for the monopole of the halo--matter cross power spectrum $P_{hm}^{\ell=0}(k)$; halos in redshift space, matter field in real space. The bias parameters used for predictions are the ones inferred from the $k$NN-CDFs. The grey shaded region represents the scatter in the measurements across 100 realisations. The agreement here is limited to $k=0.25\,h/\mathrm{Mpc}$, similar to the galaxy case.}
    \label{fig:RsdHalos_Pgm0_1024p_Resid}
\end{figure}

We also validate our modelling pipeline based on the hybrid Lagrangian bias expansion on dark matter halos taken from the high-resolution \textit{Quijote} simulations at redshift $z=0.5$ (the same suite used for the galaxy sample defined in Section~\ref{sec:sim_tracers}). We work with $2\times10^5$ halos in the mass range $10^{13}-10^{13.5}\,\mathrm{M}_{\odot}/h$. For the $k$NN-CDF measurements, we further subsample this halo sample twice to probe the scales of interest ($10\,\mathrm{Mpc}/h$ to $30\,\mathrm{Mpc}/h$), resulting in an effective number density of $\bar{n}_{\rm eff}=10^{-4}\,(h/\mathrm{Mpc})^3$. The halo sample consists of central halos only; thus, while modelling the RSDs, we need to account for only the central distortion. 

The residuals between the $k$NN-CDF measurements and the corresponding model predictions are presented in Figures~\ref{fig:RealHalos_knnCdf_1024p_Resid} and \ref{fig:RsdHalos_knnCdf_1024p_Resid} for real and redshift space, respectively. The figure conventions are the same as in our main results for galaxies (Section~\ref{sec:ModellingResults}). In real space, the 1NN-CDF is the best-modelled case, with good agreement extending to $r=12\,\mathrm{Mpc}/h$. In redshift space, the model validity is limited on small scales; the 1NN-CDF remains the best-modelled case, with good agreement extending to $r=16\,\mathrm{Mpc}/h$ for both auto and cross clustering.

The bias parameters inferred from the halo $k$NN-CDFs also provide a good description of the halo power spectrum. The residuals between $P(k)$ measurements and model predictions are shown in Figures~\ref{fig:RealHalos_Pk_1024p_Resid} and \ref{fig:RsdHalos_Pk_1024p_Resid} for real and redshift space, respectively. The figure layout and pattern are the same as in our main results (Section~\ref{sec:ModellingResults}). While the monopole is well described down to $k=0.6\,h/\mathrm{Mpc}$, agreement for the quadrupole is limited to larger scales ($k \leq 0.25 \, h/\mathrm{Mpc}$), indicating a limitation of the current redshift-space model for halos. This limitation is not seen in the corresponding galaxy analysis, possibly because the two FoG parameters provide greater flexibility in describing redshift-space clustering. We also present the halo--matter cross power spectrum (halos in redshift space and matter field in real space) in Figure~\ref{fig:RsdHalos_Pgm0_1024p_Resid}. The model describes the measurements up to $k=0.25\,h/\mathrm{Mpc}$, similar to what was obtained for galaxies.

\section{Supplementary Results for the Fisher Analysis}
\label{sec:Fisher_Appendix}
\begin{figure}
    \centering
    \includegraphics[width=0.625\linewidth]{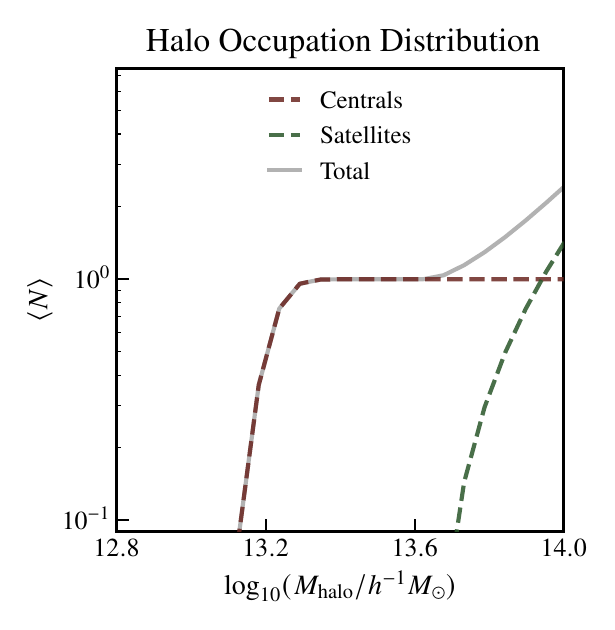}
    \caption{The Halo Occupation Distribution (HOD) of the galaxies used in our Fisher analysis. The galaxies are generated by populating the halo catalogues at $z_f=0.5$, showing the central (maroon dashed), satellite (green dashed), and total (grey solid) galaxy occupations as a function of halo mass. The HOD parameters adopted for this analysis are $\log M_{\rm min}=13.2$, $\sigma_{\log M}=0.075$, $\log M_0=13.64$, $\log M_1=13.64$, and $\alpha=1.35$. Although this HOD is not intended to represent a very realistic galaxy sample, it provides a controlled setup for carrying out the Fisher forecasts.}
    \label{fig:GalxHOD_512p}
\end{figure}

We performed the Fisher analysis based on galaxy auto clustering and galaxy--matter cross clustering (galaxies in redshift space, matter field in real space; HOD for galaxy sample shown in Figure~\ref{fig:GalxHOD_512p}). The bias parameters inferred from the $k$NN-CDFs of these galaxies also provide a good fit to the $P(k)$ measurements in redshift space when combined with the FoG parameters. We therefore used these bias and FoG parameters as the fiducial set for the Fisher analysis. Here, we first show the residuals between the measurements and model predictions obtained with this fiducial set for the $k$NN-CDFs and $P(k)$, followed by additional Fisher forecast results.

\begin{figure*}
    \centering
    \includegraphics[width=0.85\linewidth]{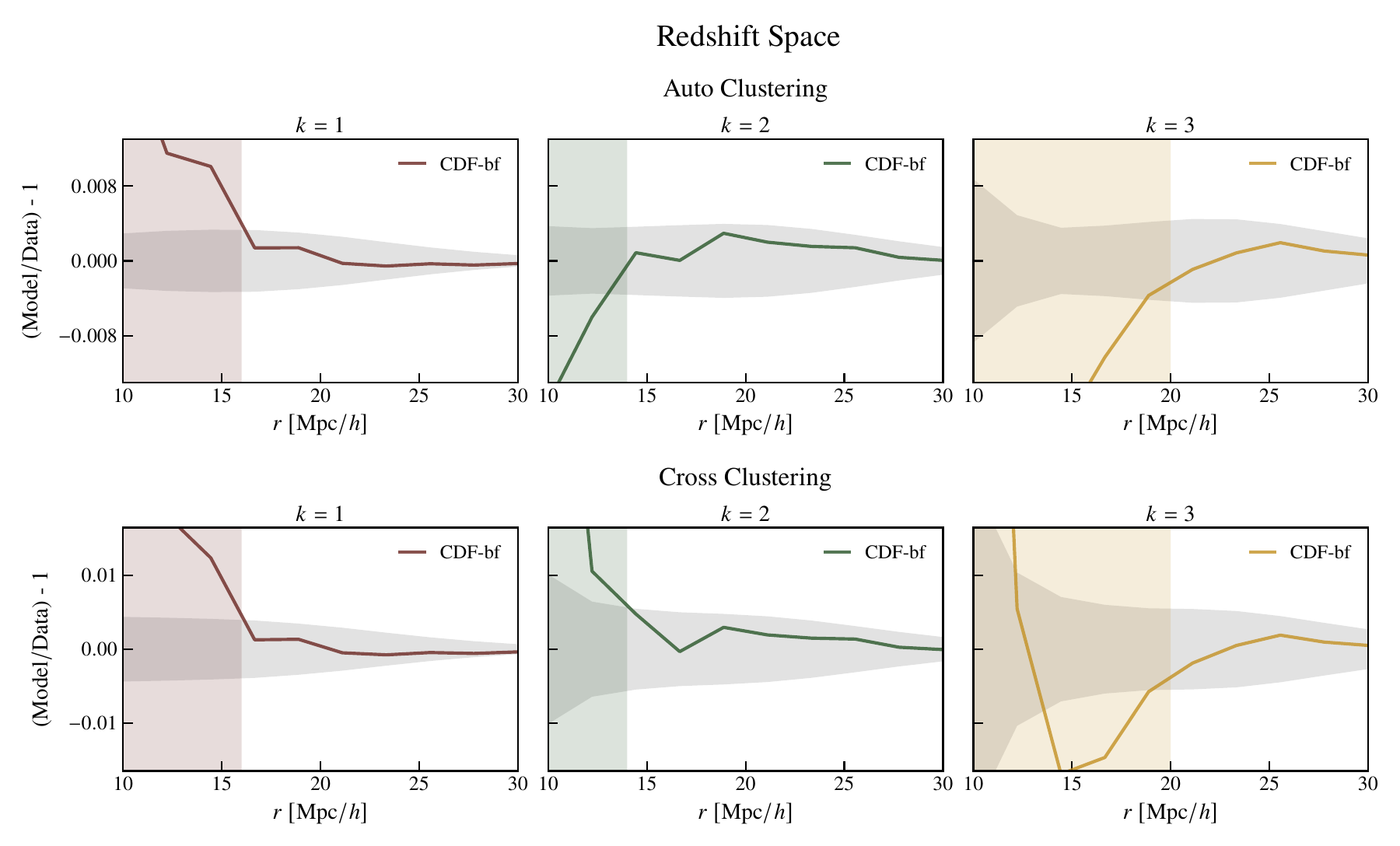}
    \caption{Redshift space residuals between $k$NN-CDF measurements on galaxies and the corresponding model predictions within the hybrid Lagrangian bias framework, for the first three nearest-neighbours ($k=1,2,3$). The top row presents the residuals for galaxy auto clustering, while the bottom row presents the same for galaxy--matter cross clustering (matter field in real space). The grey shaded region in each panel denotes the scatter in measurements across 100 realisations. The shaded region extending up to a given $r$ in each panel marks the scales where the agreement between data and model does not hold. The agreement between data and model extends down to $16\,\mathrm{Mpc}/h$, $14\,\mathrm{Mpc}/h$ and $20\,\mathrm{Mpc}/h$ for the 1NN-, 2NN- and 3NN-CDFs, respectively. We restrict our Fisher analysis to the scales where agreement holds.}
    \label{fig:RsdAutoCross_knnCdf_512p_Resid}
\end{figure*}

\begin{figure}
    \centering
    \includegraphics[width=0.975\linewidth]{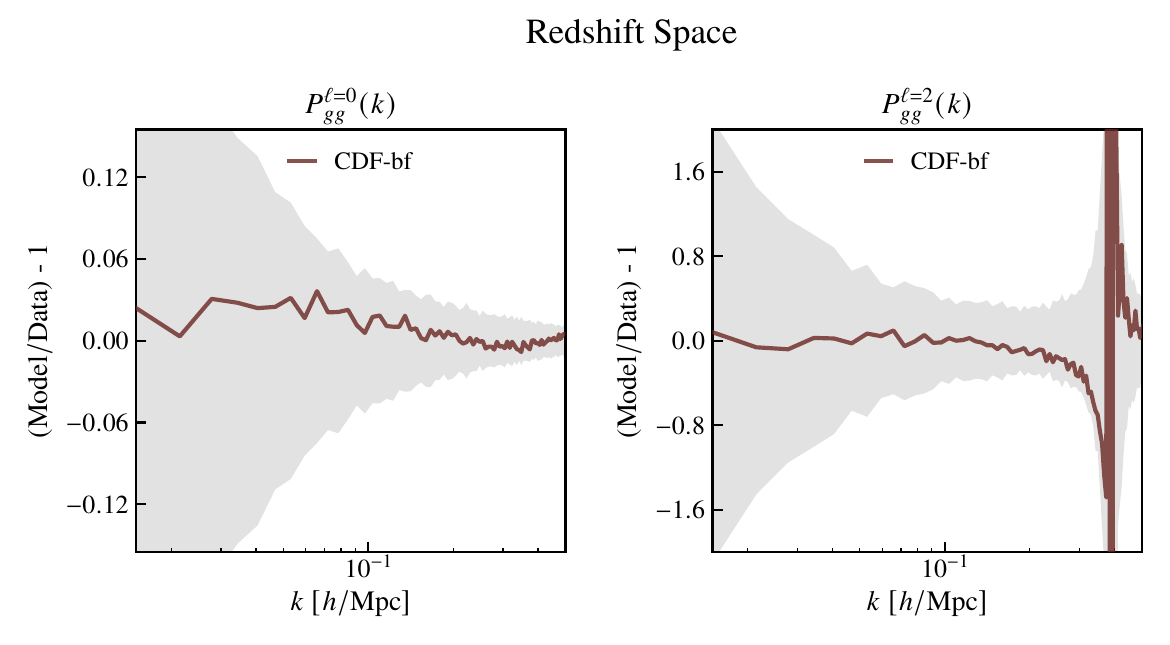}
    \caption{Residuals for $P_{gg}^{\ell=0}(k)$ and $P_{gg}^{\ell=2}(k)$ between the galaxy measurements in redshift space and the corresponding predictions obtained within the hybrid Lagrangian bias framework using the model parameters inferred from the $k$NN-CDFs. The grey shaded region denotes the scatter in measurements across 100 realisations. The $k$NN-CDF inferred parameters can describe the $P(k)$ measurements up to $k= 0.5 \, h/\mathrm{Mpc}$, allowing the two statistics to be combined for Fisher forecasts based on a common parameter set.}
    \label{fig:RsdAuto_Pk_512p_Resid}
\end{figure}

\begin{figure}
    \centering
    \includegraphics[width=0.525\linewidth]{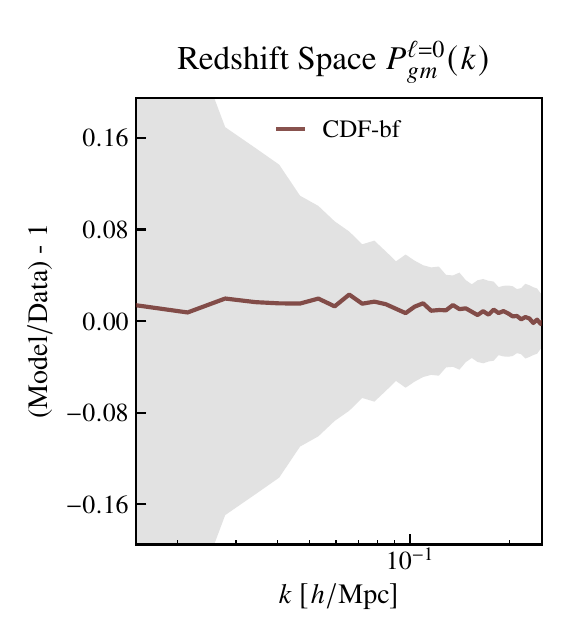}
    \caption{Residuals for $P_{gm}^{\ell=0}(k)$ between galaxy measurements and predictions obtained within the hybrid Lagrangian bias framework using the bias and FoG parameters inferred from the $k$NN-CDFs. Here, the galaxies are in redshift space while the matter field is in real space. The grey shaded region denotes the scatter in the measurements across 100 realisations. The $k$NN-CDF inferred bias parameters can describe the $P_{gm}^{\ell=0}(k)$ data only up to $k = 0.25\,h/\mathrm{Mpc}$; thus we use this range for the Fisher forecasts based on a common set of best-fit bias parameters.}
    \label{fig:RsdPgm0_512p_Resid}
\end{figure}

Figure~\ref{fig:RsdAutoCross_knnCdf_512p_Resid} shows the residuals between the measured and predicted redshift-space $k$NN-CDFs. The top row shows galaxy auto clustering, while the bottom row shows galaxy--matter cross clustering, with galaxies in redshift space and the matter field in real space. The agreement between data and model extends down to $16\,\mathrm{Mpc}/h$, $14\,\mathrm{Mpc}/h$ and $20\,\mathrm{Mpc}/h$ for the 1NN-, 2NN- and 3NN-CDFs, respectively. These are the scales used for the respective $k$NN-CDFs in the Fisher analysis.
The corresponding residuals for the redshift-space $P(k)$ measurements are shown in Figure~\ref{fig:RsdAuto_Pk_512p_Resid}. The left and right panels show $P_{gg}^{\ell=0}(k)$ and $P_{gg}^{\ell=2}(k)$, respectively. The predictions obtained with the fiducial parameter set describe the measurements well up to $k=0.5\,h/\mathrm{Mpc}$. We therefore use $0.015\,h/\mathrm{Mpc} \leq k \leq 0.5\,h/\mathrm{Mpc}$ in the $P(k)$ auto clustering Fisher analysis. Figure~\ref{fig:RsdPgm0_512p_Resid} shows the residuals for $P_{gm}^{\ell=0}(k)$, with galaxies in redshift space and the matter field in real space. The model describes these measurements well up to $k=0.25\,h/\mathrm{Mpc}$, and we use this range for the Fisher analysis.


Here, we present additional comparisons of Fisher forecasts from the same pipeline.

\begin{figure*}
    \centering
    \includegraphics[width=0.95\linewidth]{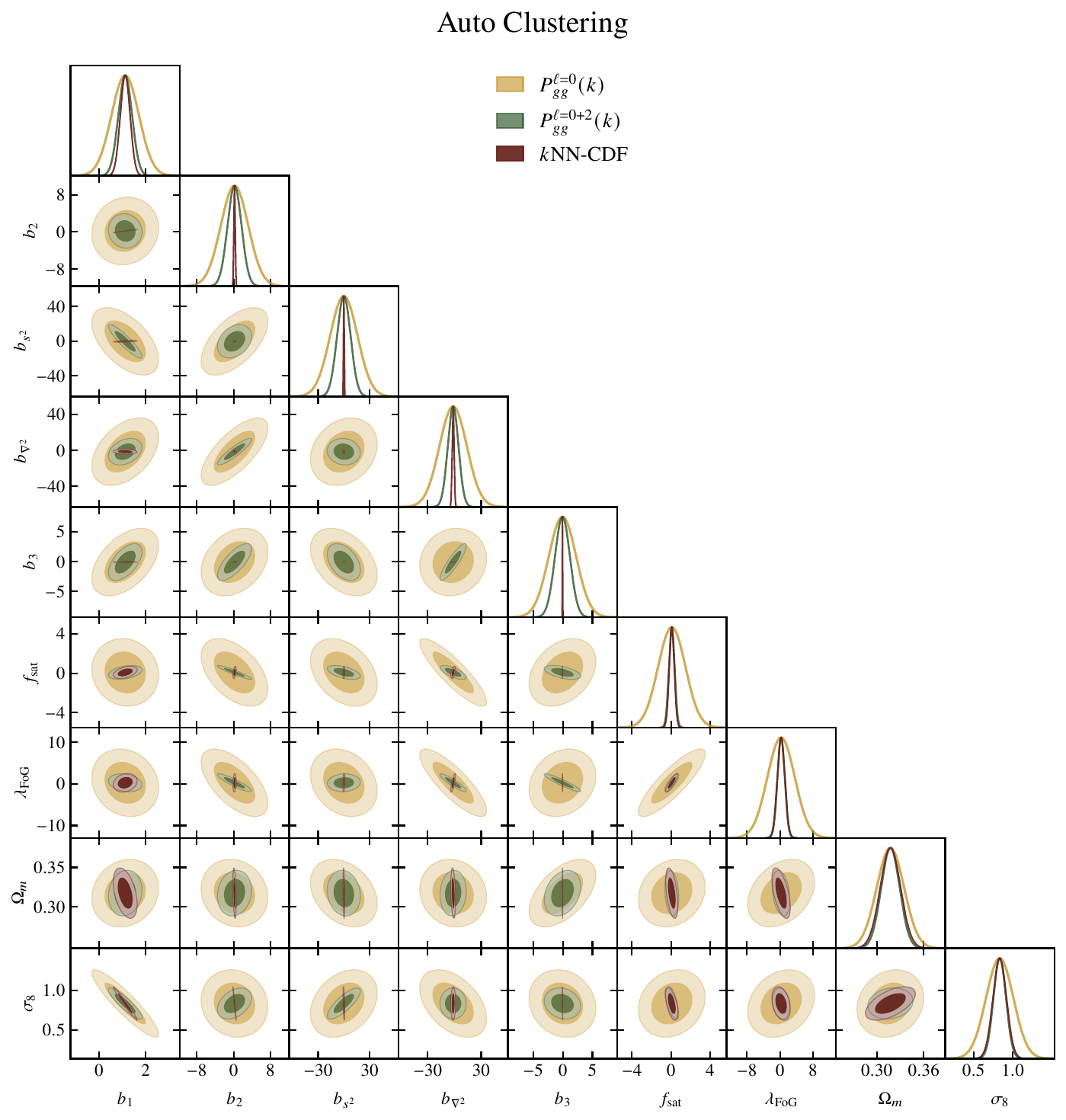}
   \caption{Fisher constraints on the bias ($b_1$, $b_2$, $b_{s^2}$, $b_{\nabla^2}$, $b_3$), FoG ($f_{\rm sat}$, $\lambda_{\rm FoG}$), and cosmological ($\Omega_m$ and $\sigma_8$) parameters, from galaxy auto clustering using $P_{gg}^{\ell=0}(k)$ , $P_{gg}^{\ell=0+2}(k)$, and the $k$NN-CDFs ($k=1,2,3$). Off-diagonal panels show the marginalised $1\sigma$ and $2\sigma$ two-dimensional posterior contours, while diagonal panels show the marginalised one-dimensional posteriors. Including the quadrupole improves the $P(k)$ constraints on all parameters, and more significantly for $f_{\rm sat}$ and $\lambda_{\rm FoG}$. Comparing constraints from $P_{gg}^{\ell=0+2}(k)$ and the $k$NN-CDFs, the latter yields tighter constraints on all bias parameters, more so for the higher-order biases, reflecting their sensitivity to higher-order clustering information that $P(k)$ does not capture. For the remaining parameters, the constraints from $P(k)$ and the $k$NN-CDFs are comparable.}
    \label{fig:Fisher_AutoClstr_Apndx}
\end{figure*}

In Figure~\ref{fig:Fisher_AutoClstr_Apndx}, we compare Fisher constraints on the bias ($b_1$, $b_2$, $b_{s^2}$, $b_{\nabla^2}$, $b_3$), FoG ($f_{\rm sat}$, $\lambda_{\rm FoG}$), and cosmological ($\Omega_m$ and $\sigma_8$) parameters, from galaxy auto clustering using the $P(k)$ monopole ($\ell=0$), $P(k)$ monopole and quadrupole ($\ell=0+2$), and the $k$NN-CDFs. Including the quadrupole improves the $P(k)$ constraints on all parameters, particularly $f_{\rm sat}$ and $\lambda_{\rm FoG}$. Comparing constraints from $P_{gg}^{\ell=0+2}(k)$ and the $k$NN-CDFs, the latter yields tighter constraints on all bias parameters, with a particularly great improvement for the higher-order biases, reflecting their sensitivity to higher-order clustering information that $P(k)$ does not capture. For the remaining parameters, the constraints from $P(k)$ and the $k$NN-CDFs are comparable.

\begin{figure*}
    \centering
    \includegraphics[width=0.95\linewidth]{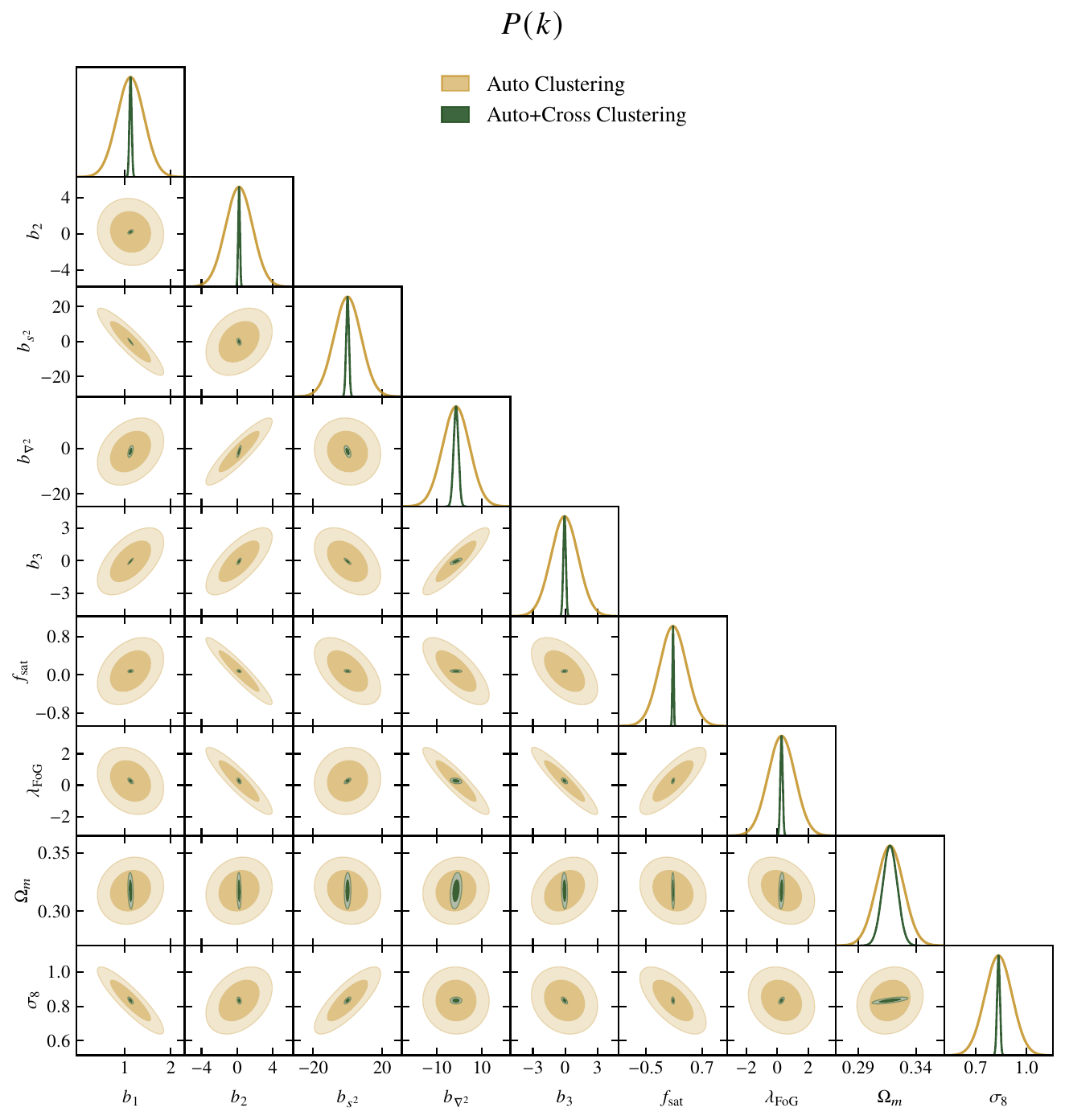}
    \caption{Fisher constraints on the bias ($b_1$, $b_2$, $b_{s^2}$, $b_{\nabla^2}$, $b_3$), FoG ($f_{\rm sat}$, $\lambda_{\rm FoG}$), and cosmological ($\Omega_m$ and $\sigma_8$) parameters, obtained from the $P(k)$ analysis for galaxies in redshift space, comparing galaxy auto clustering with the combination of galaxy auto and galaxy--matter cross clustering (matter field in real space). Off-diagonal panels show the marginalised $1\sigma$ and $2\sigma$ two-dimensional posterior contours, while diagonal panels show the marginalised one-dimensional posteriors. The constraints in golden yellow correspond to auto clustering case, whereas those in green correspond to the combined case. The scales used here are the same as used in the main Fisher analysis. The largest improvement by using the combination is seen for $f_{\rm sat}$, while the constraints on $\sigma_8$ and $\Omega_m$ improve by factors of about 10 and 1.8, respectively.}
\label{fig:Fisher_Pk_AutoCrossClstr_Apndx}
\end{figure*}

\begin{figure*}
    \centering
    \includegraphics[width=0.95\linewidth]{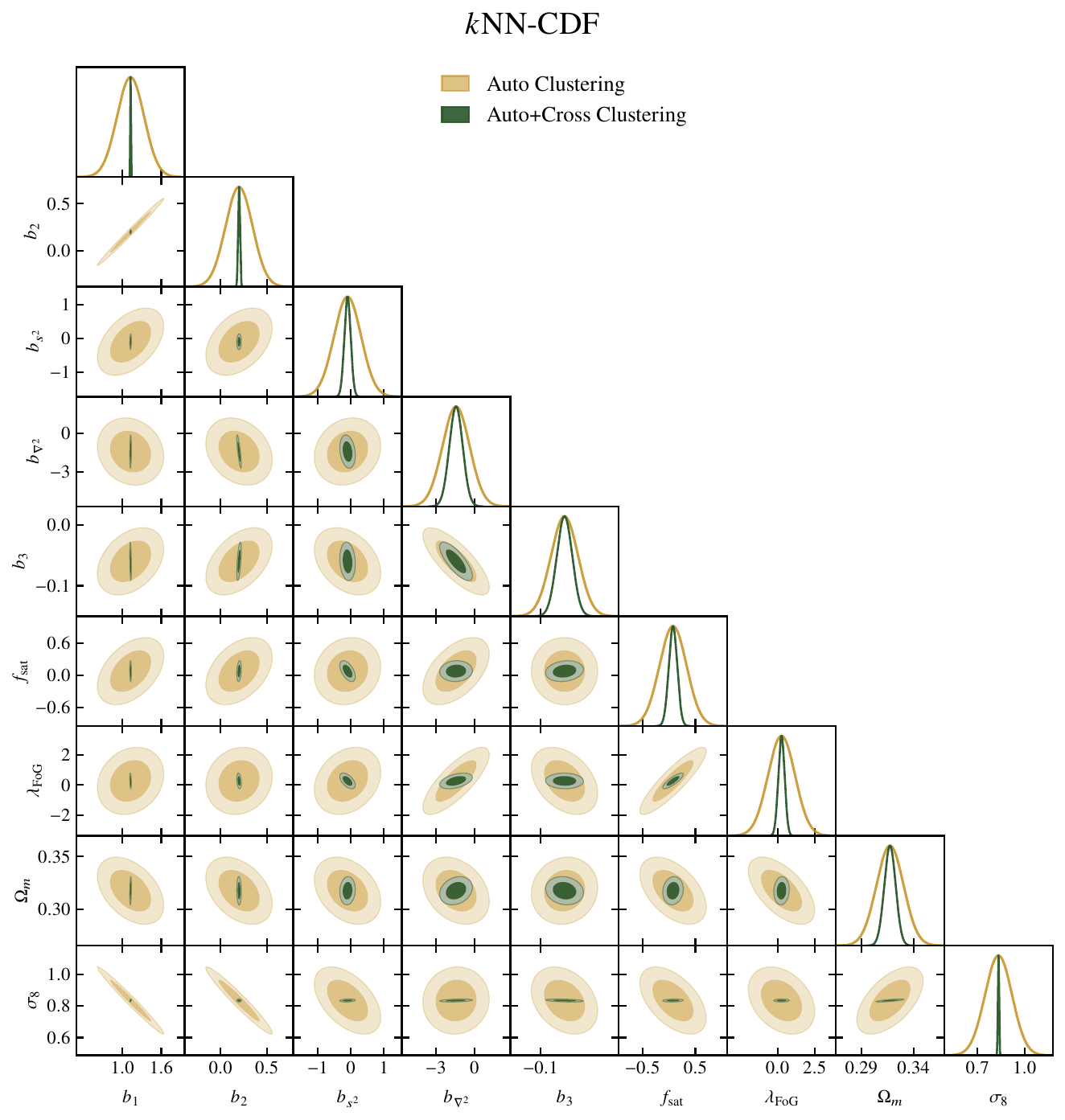}
   \caption{Fisher constraints on the bias ($b_1$, $b_2$, $b_{s^2}$, $b_{\nabla^2}$, $b_3$), FoG ($f_{\rm sat}$, $\lambda_{\rm FoG}$), and cosmological ($\Omega_m$ and $\sigma_8$) parameters, obtained from $k$NN-CDFs ($k=1,2,3$) of galaxies in redshift space, comparing galaxy auto clustering with the combination of galaxy auto and galaxy--matter cross clustering (matter field in real space). Off-diagonal panels show the marginalised $1\sigma$ and $2\sigma$ two-dimensional posterior contours, while diagonal panels show the marginalised one-dimensional posteriors. We use the same scales and nearest-neighbour orders as in the main Fisher analysis. The constraints in golden yellow correspond to auto clustering case, whereas the ones in green correspond to the combined case. Including cross clustering information significantly improves the constraints on all parameters, with the largest gain for $b_1$, while the constraints on $\sigma_8$ and $\Omega_m$ improve by factors of about 23 and 2.4, respectively.}
\label{fig:Fisher_knnCdf_AutoCrossClstr_Apndx}
\end{figure*}

Figure~\ref{fig:Fisher_Pk_AutoCrossClstr_Apndx} compares the Fisher constraints from the redshift-space galaxy auto power spectrum (monopole and quadrupole) with those from combining the galaxy auto and galaxy--matter cross power spectra. For the cross spectrum, galaxies are in redshift space and the matter field is in real space, as described in the main text. We use the same scales as in the main Fisher analysis. Adding the cross spectrum improves the constraints on all parameters, with the largest improvement for $f_{\rm sat}$, while the constraints on $\sigma_8$ and $\Omega_m$ improve by factors of about 10 and 1.8, respectively. This demonstrates the additional constraining power of galaxy--matter cross clustering when combined with galaxy auto clustering.
Figure~\ref{fig:Fisher_knnCdf_AutoCrossClstr_Apndx} shows the corresponding improvement when galaxy--matter cross clustering information is added to the galaxy auto clustering $k$NN-CDFs. The constraints improve significantly for all parameters. We use the same scales and nearest-neighbour orders as in the main Fisher analysis. The largest improvement is for the linear bias parameter $b_1$, while the constraints on $\sigma_8$ and $\Omega_m$ improve by factors of about 23 and 2.4, respectively. These results demonstrate the additional constraining power of galaxy--matter cross clustering when combined with galaxy auto clustering $k$NN-CDFs.


\bsp	
\label{lastpage}
\end{document}